\documentclass[apj]{emulateapj}
\usepackage{bm}
\usepackage{multirow}
\usepackage{color}

\usepackage{amsmath}
\usepackage{amsfonts}
\usepackage{amssymb}
\usepackage{epsfig}
\usepackage{graphicx}
\usepackage{hyperref}

\newcommand{\be}{\begin{equation}}
\newcommand{\ee}{\end{equation}}
\newcommand{\ba}{\begin{eqnarray}}
\newcommand{\ea}{\end{eqnarray}}

\newcommand{\beq}{\begin{equation}}
\newcommand{\eeq}{\end{equation}}
\newcommand{\beqa}{\begin{eqnarray}}
\newcommand{\eeqa}{\end{eqnarray}}

\begin{document}

\title{Cross-Correlation of Near and Far-Infrared Background Anisotropies as Traced by {\it Spitzer} and {\it Herschel}}
\author{Cameron Thacker$^{1}$, Yan Gong$^{1}$, Asantha Cooray$^{1}$, Francesco De Bernardis$^{2}$,
Joseph Smidt$^{3}$, Ketron Mitchell-Wynne$^{1}$}

\affiliation{$^{1}$Department of Physics  and  Astronomy, University  of
California, Irvine, CA 92697 USA}
\affiliation{$^{2}$Department of Physics, Cornell University, Ithaca, NY 14853 USA}
\affiliation{$^{3}$T-2, Los Alamos National Laboratory, Los Alamos, NM 87545 USA}

\begin{abstract}
    We present the cross-correlation between the far-infrared background fluctuations as measured
    with the {\it Herschel} Space Observatory at 250, 350, and 500 $\mu$m and the near-infrared
    background fluctuations with {\it Spitzer} Space Telescope at 3.6 $\mu$m. The cross-correlation
    between far and near-IR background anisotropies are detected such that the correlation
    coefficient at a few to ten arcminute angular scales decreases from 0.3 to 0.1 when the far-IR
    wavelength increases from 250 $\mu$m to 500 $\mu$m. We model the cross-correlation using a halo
    model with three components: (a) far-IR bright or dusty star-forming galaxies below the masking
    depth in {\it Herschel} maps, (b) near-IR faint galaxies below the masking depth at 3.6 $\mu$m,
    and (c) intra-halo light, or diffuse stars in dark matter halos, that is likely dominating
    fluctuations at 3.6 $\mu$m. The model is able to reasonably reproduce the auto correlations at
    each of the far-IR wavelengths and at 3.6 $\mu$m and their corresponding cross-correlations.
    While the far and near-IR auto-correlations are dominated by faint dusty, starforming galaxies
    and intra-halo light, respectively, we find that roughly half of the cross-correlation between
    near and far-IR backgrounds is due to the same dusty galaxies that remain unmasked at 3.6
    $\mu$m.  The remaining signal in the cross-correlation is due to intra-halo light present in the
    same dark matter halos as those hosting the same faint and unmasked galaxies.  In this model the
    decrease in the cross-correlation signal from 250 $\mu$m to 500 $\mu$m comes from the fact that
    the galaxies that are primarily contributing to 500 $\mu$m fluctuations peak at a higher
    redshift than those at 250 $\mu$m.


\end{abstract} 

\maketitle

%%%
\section{Introduction}

The cosmic infrared background (CIB) contains the total emission history of the Universe integrated
along the line of sight. The CIB contains two peaks, one at optical/near-IR wavelengths around 1
$\mu$m and the second at far-IR wavelengths around 250 $\mu$m  \citep{Dole2006}. The former is
composed of photons produced during nucleosynthesis in stars while the latter is reprocessing of
some of those photons by dust in the universe. While the total intensity in the near-IR background,
especially at 3.6 $\mu$m, has been mostly resolved to individual galaxies, we are still far from
directly resolving the total CIB intensity at 250 $\mu$m and above to individual sources. This is
due to the fact that at far-IR wavelengths observations are strongly limited by the aperture sizes.
With the SPIRE Instrument \citep{Griffin:2010hp} aboard the {\it Herschel} Space
Observatory \citep{Pillbratt2010} \footnote{Herschel is an ESA space observatory with science instruments provided by
European-led Principal Investigator consortia and with important participation from NASA.}, the background has been resolved to 5\%, 15\% and 22\% at  250 $\mu$m, 350
$\mu$m and 500 $\mu$m, respectively \citep{Oliver2010}. In order to understand some properties of
the faint far-IR sources it is essential that we study the fluctuations or the anisotropies of the
background.

These spatial fluctuations in the CIB are best studied using the angular power spectrum. This
technique provides a way to study the faint and unresolved galaxies because while not individually
detected, they trace the large scale structure. The clustering of these galaxies is then measurable
through the angular power spectrum of the background intensity variations \citet{Amblard:2011gc}.
Such fluctuation clustering measurements at far-IR wavelengths have been followed up with both {\it
Herschel} and Planck \citep{Viero2013,Planck2014}, the latter of which has provided the highest
signal to noise calculation in the far-IR.

Separately near-IR background anisotropies have been studied in the literature and have been
interpreted as  due to galaxies containing PopIII stars present during reionization
\citep{Kashlinsky2005,Kashlinsky2007,Kashlinsky2012}, direct collapse black holes at $z > 12$
\citep{Yue2013}, and intra-halo light \citep{Cooray2012,Zemcov2014}.  In addition to {\it Spitzer} at 3.6
$\mu$m and above, fluctuation measurements in the near-IR wavelengths have come from Akari
\citep{Matsumoto2011} and recently with the Cosmic Infrared Background Experiment (CIBER) at 1.1 and
1.6 $\mu$m \citep{Zemcov2014}. While earlier studies argued for a substantial contribution from
galaxies at $z > 8$, including those with PopIII stars or blackholes, recent studies find
that such signals are not likely to be the dominant contribution \citep{Cooray2012,Yue2014}. Both
\citet{Cooray2012} using {\it Spitzer} and \citet{Zemcov2014} using CIBER argue for the case that the
signal is coming from low redshifts and proposes an origin that may be associated with intra-halo
light.  Intra-halo light invovles diffuse stars that are tidally stripped during galaxy mergers and
other interactions.  However, we still expect some signal from galaxies during reionization. In
addition to these two, there is also a third contribution. These are the faint, dwarf galaxies that
are present between us and reionization. Such galaxies contribute to the near-IR and since they have
flux densities below the point source detection level in near-IR maps, they remain unmasked during
fluctuation power spectrum measurements. The exact relative amplitudes of each of these three
signals are yet to be determined.

While fluctuations have been studied separately at far-IR and near-IR wavelengths, no attempt has
been made to combine those measurements yet. In this work, we present the first results from just
such a cross-correlation. We make use of the overlap coverage in the eight square degrees Bo\"otes
field between SDWFS at 3.6 $\mu$m \citep{Ashby2009} and {\it Herschel} at 250, 350 and 500 microns. We
find that the two signals are correlated but the cross-correlation coefficient is weak at 30\% or
below. Such a weak cross-correlation argues for the scenario that {\it Spitzer} and {\it Herschel} are mostly
tracing two differerent populations. To interpret the data, we model the auto and cross-correlation
signals using a three component halo model composed of far infrared (FIR) galaxies, intra-halo light (IHL), and faint galaxies.

The paper is structured as follows. In Section 2, we discuss the data analysis and power spectra
measurements. This includes the map making process for the {\it Herschel} and {\it Spitzer} data using
HIPE and Self-Calibration, respectively. This section also explains the mask generation procedure, defines the
cross-correlation power spectrum, and discusses sources of error. In Section 3, we describe the halo
model including components for FIR galaxies, IHL, and faint galaxies and the MCMC process used to
fit the data. Finally, in Sections 4 and 5 we present the results of our model fit, discuss the
their implications, and give our concluding thoughts.

%%%%

\section{Data Analysis and Power Spectra Measurements}

We discuss the analysis pipeline we implemented for the cross-correlation of {\it Herschel} and
{\it Spitzer} fluctuations. The study is done in the Bo\"otes field making use of the 
{\it Herschel}/SPIRE instrument 
data taken as part of the {\it Herschel} Multi-tiered Extragalactic Survey (HerMES; Oliver et al.
2012) and {\it Spizter}/IRAC imaging data taken as part of the {\it Spitzer} Deep Wide Field Survey (SDWFS) \citep{Ashby2009}.
We describe the datasets, map-making, source masking and power spectrum measurement details in the
sub-sections below.

\subsection{Map Making}

\subsubsection{Far-IR maps with Herschel}

We make use of the publicly available SPIRE instrument data of the Bo\"otes field taken as part of
HerMES from the ESA {\it Herschel} Science Data Archive\footnote{Observation Id's listed in
acknowledgements}. Our map-making pipeline makes use of the Level 1 time-ordered scans that have been
corrected for cosmic rays, temperature drifts, and bolometer time reponse as part of the standard
data reduction by ESA.  In those timelines we remove a baseline polynomial in a scan-by-scan basis
to normalize gain variations.  The SPIRE maps are generated using the {\it Herschel} Interactive
Processing Environment (HIPE) \citep{Ott2010}. For this work we make use of the MADmap
\citep{Cantalupo2010} algorithm that is native to HIPE. It is a maximum likelihood estimate and
follows the approach used in \citet{Thacker2013}. We do not use the map-maker that was developed
for anisotropy power spectrum measurement in \citep{Amblard2010} due to improvements that have
been made to the HIPE and its bulit-in map-makers since the Amblard et al. study in 2010.

The end products of the map-making process are  single tile maps covering roughly 8 square degrees
and containing two sets of 80 scans in orthogonal directions with a pixel scale of 6 arcseconds per
pixel in all SPIRE bands (250$\mu$m, 350$\mu$m, and 500$\mu$m). These maps, using publicly available
data from HerMES, are generated at a pixel scale one third of the beam size consistent with our
previous work as well.  Such a sampling allows the point source fluxes to be measured more
accurately with an adequate sampling of the PSF.  For this study, we are forced to compromise
between an accurate sampling of the PSF, point source detection and flux density measurement to get an
accurate combination of {\it Herschel} and {\it Spitzer} data. Since the native scale of  {\it
Spitzer} maps are at 1.2 arcseconds/pixel, we do not want to increase the SPIRE pixel scale
substantially. We will return to this issue more below when we dicuss our point source masks that
were applied to the maps to remove detected sources.

\subsubsection{{\it Spitzer}}
                                                                        
The SDWFS \citep{Ashby2009} maps of the Bo\"otes field consist of four observations taken over 7 to
10 days from 2004 to 2008\footnote{http:sha.ipac.caltech.edu/applications/Spitzer/SHA/ \\ Program
number GO40839 (PI. D.  Stern)}. For this analysis we limit our data to the 3.6$\mu$m channel. With
four sets of data taken at different roll angles we can ensure our fluctuation measurements are
robust to detector systematics.  In addition, since the measurements were taken in different years
and months, multiple jack-knife tests enable us to reduce and quantify the error from astrophysical
systematics such as the zodiacal light (ZL). 

To further control the errors and to obtain a uniform background measurement across a very wide area
the maps are mosaiced using a {\sc Self-Calibration} algorithm \citep{Arendt2000}. The algorithm is able
to match the sky background levels with free gain parameters between adjacent overlapping frames
using a least squares fitting technique. We make use of the cleaned basic calibrated data (cBCDs)
publicly available from the {\it Spitzer} data archive for the map-making process. These frames have
asteroid trails, hot pixels, and other artifacts removed. Our original maps are made at a pixel
scale of  1.2 arcsecond per pixel.  Unlike our previous work \citep{Cooray2012}, we have repixelized
the maps to a 6 arcsecond per pixel scale to match with the pixel scale of our {\it Herschel} maps.

\begin{figure}
    \begin{center}
    \includegraphics[scale=.5,clip]{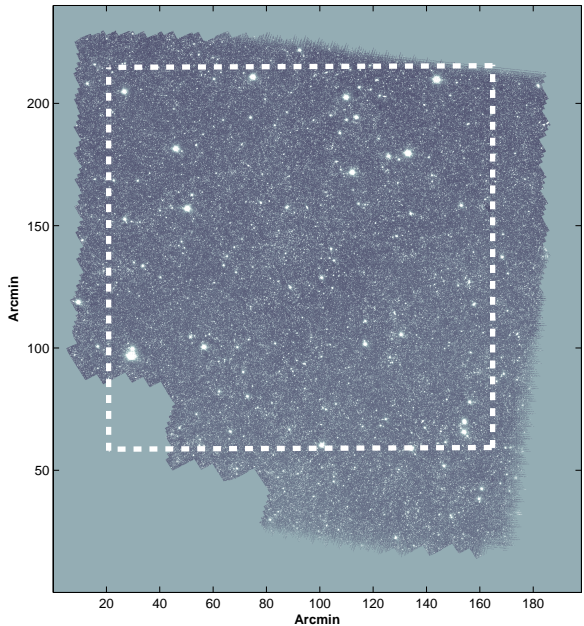}
    \end{center}
    \caption{{\it Spitzer} 3.6$\mu$m Bo\"otes field with only the overlapping area with {\it Herschel}. 
The white dashed lines show the cropped region used for the cross-correlation study. We crop to a rectangle to minimize
the edge effects and biases from the non-uniform turn-around data in the {\it Herschel}/SPIRE scan 
pattern. This region is also selected to maximize the unmasked area.}
\label{fig:crop_region}                                                                        
\end{figure}

The {\it Spitzer} maps span an area of about 10 deg$^2$, larger than the {\it Herschel}/SPIRE
coverage at 8 deg$^2$.  For this study we extract the overlapping area in both {\it Spitzer} and
{\it Herschel} as outlined in Figure \ref{fig:crop_region}.

\begin{figure*}
    \begin{center}
    \includegraphics[width=19cm]{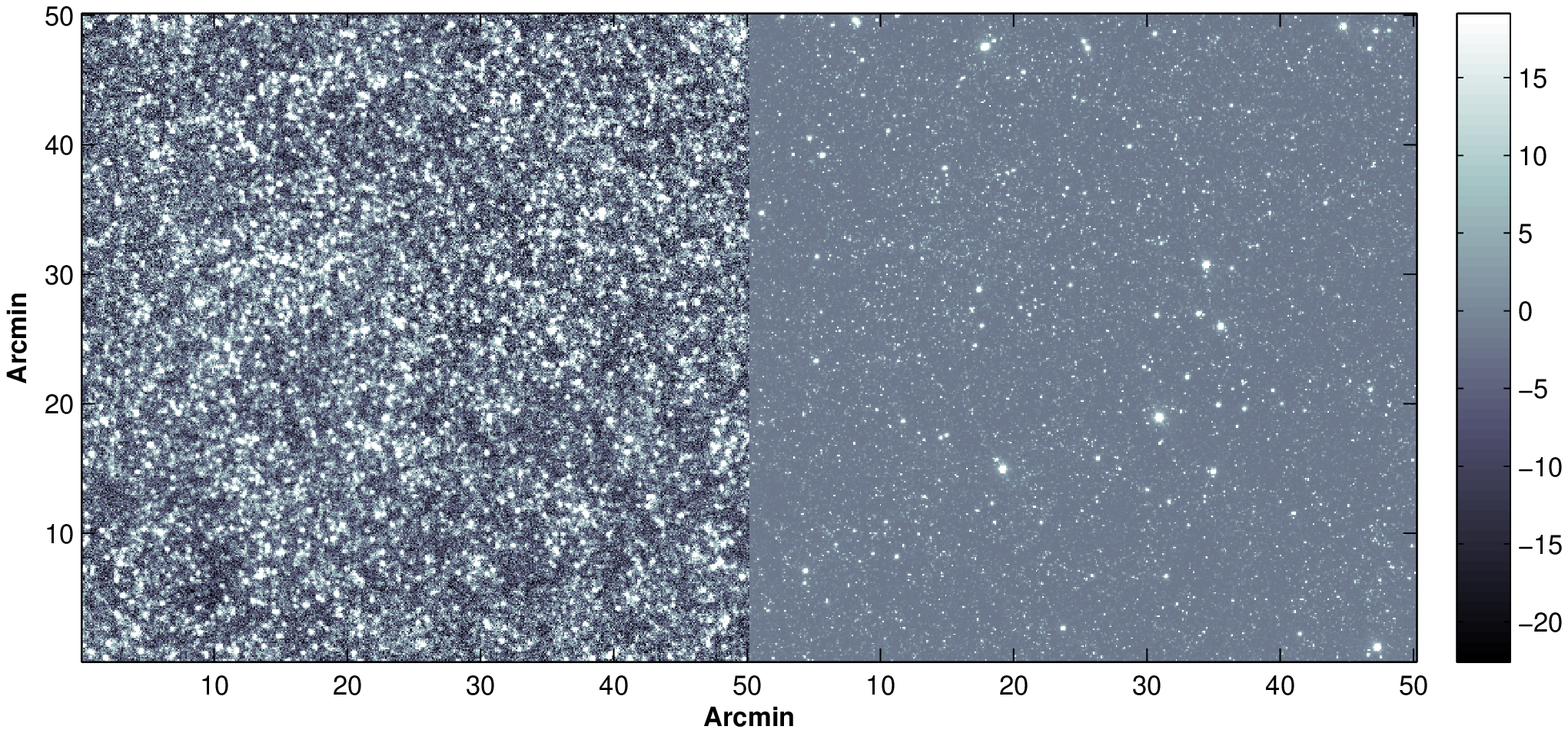}
    \includegraphics[width=19cm]{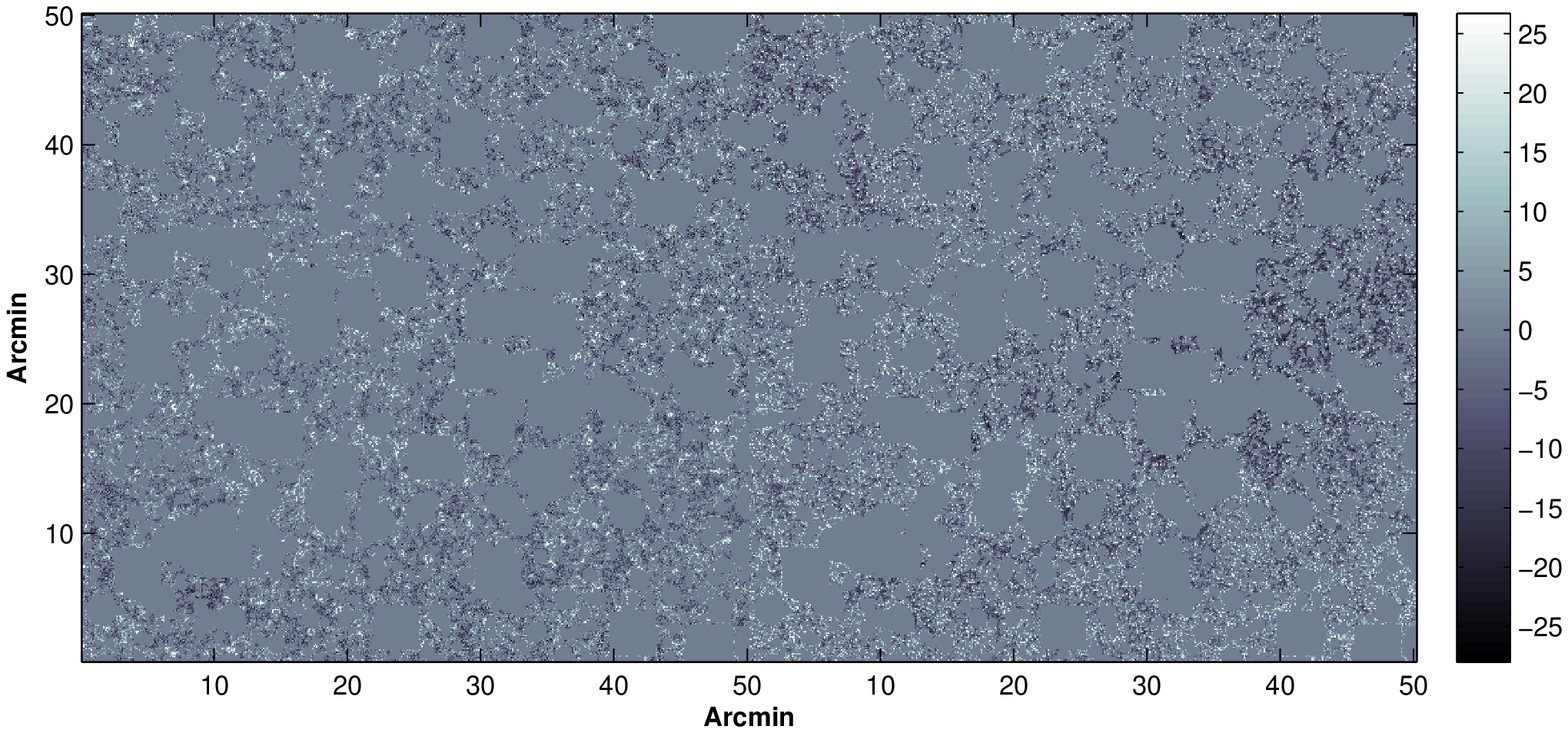}
    \end{center}
    \caption{A zoomed in image, roughly 0.7 square degrees, of  {\it Herschel} 250$\mu$m (left image) and {\it Spitzer} 3.6$\mu$m
    (right image) data of the Bo\"otes/SDWFS field. The upper images are the unmasked maps showing all sources, while the lower images have the mask applied to
remove bright detected sources from the cross-correlation study. The mask over the whole area removes 61.4\% of the pixels leaving the remainder
for the study presented here.}
      \label{fig:mask}
\end{figure*} 

\subsubsection{Astrometry}

\begin{figure}
    \begin{center}
    \includegraphics[scale=0.50,clip]{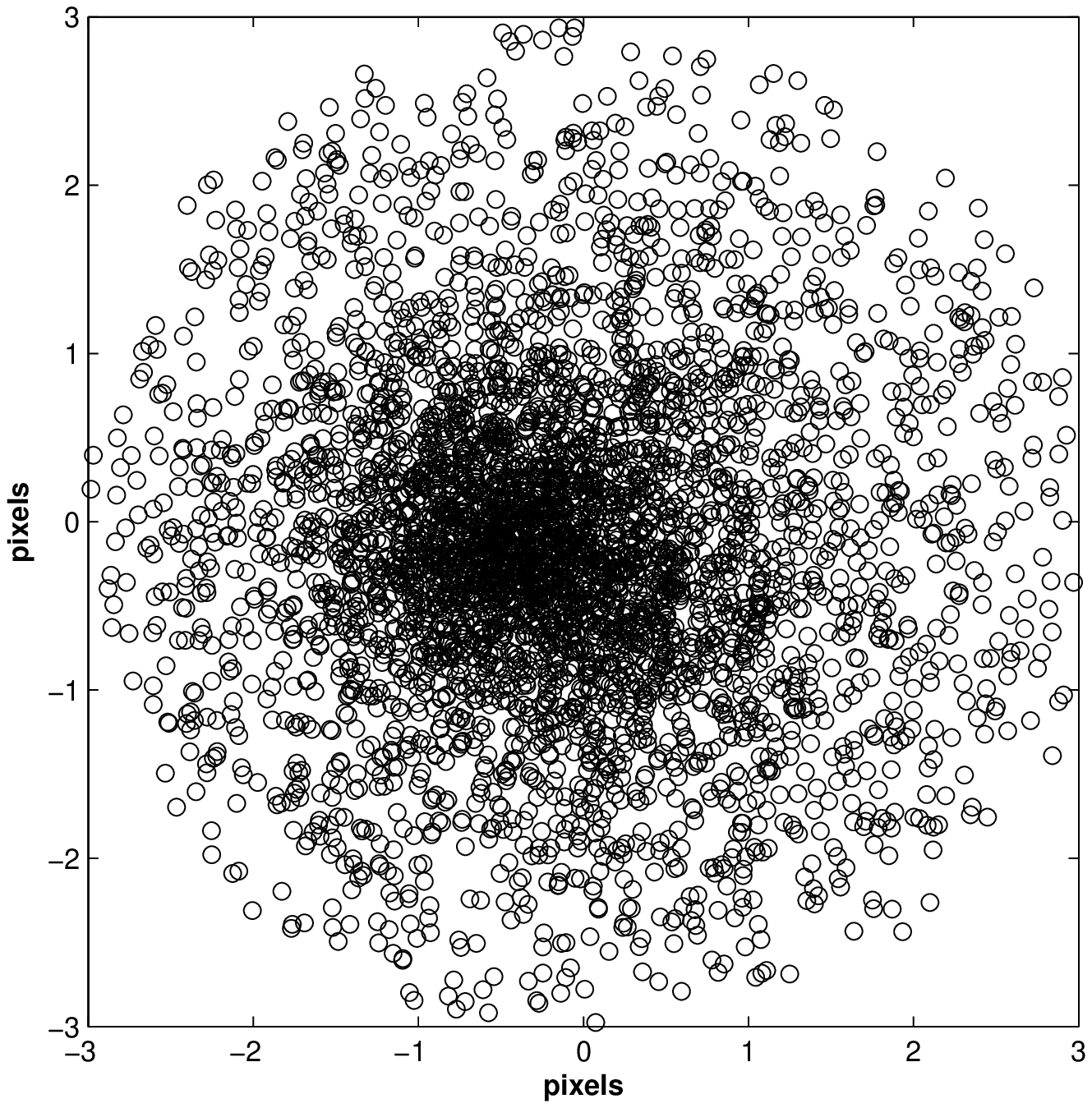}
    \caption[width=3in]{Using {\sc SExtractor} on all of the maps (the example here shows pairs from 250$\mu$m
    and 3.6 $\mu$m) we generate a list of detected sources. All of the sources within a radius of 18
    arcseconds of each map are identified to be counterparts of each other. For each counterpart we show the offset in their locations in
{\it Herschel} and {\it Spitzer} maps where we have converted differences in RA and Dec to pixels. Plotted here is the $\Delta$X and $\Delta$Y (in pixels) between
    4252 sources selected from either {\it Herschel} or {\it Spitzer} with a detected counterpart in the other map.}
\label{fig:offset}
    \end{center}
\end{figure}

To ensure both {\it Herschel} and {\it Spitzer} images are registered to the same astrometric frame,
we first checked for any offset in the {\it Spitzer} astrometry against the public catalog of SDWFS
sources and corrected the astromerty using GAIA. We then made {\it Herschel} maps to the same
astrometry as {\it Spitzer}.  As a test on our overall astrometric calibration we used Source
Extractor ({\sc SExtractor}) \citep{Bertin} to detect all sources in both the {\it Spitzer} and {\it
Herschel} images. We then matched the sources in the two catalogs within a radius of 18 arcseconds
(corresponding to the FWHM of the beam in 250$\mu$m).  Next, we took these matches and subtracted
their RA and DEC such that a perfect match would have a $\Delta$RA and $\Delta$DEC of zero.
Fig.~\ref{fig:offset} shows a scatter plot of these values in pixels rather than RA and DEC.
Perfectly aligned images would show a scatter centered at zero with equal spread in both directions.
Our analysis shows a slight offset that is less than a pixel and within our tolerance as the effect
on the cross-correlation is negligible since an offset less than a pixel will get binned to the
nearest pixel and thus produces no offest.

\label{sec:mapmaking}

%\begin{figure*}
%\begin{center}
%    \includegraphics[scale=.5,trim=0cm 0cm 0cm 0cm,clip]{spitz_fullmap.eps}
%
%    \vspace{-.708cm}
%    \includegraphics[scale=.5,trim=0cm 0cm 0cm .89cm,clip]{PSW_fullmap.eps}                                                 
%    
%    \caption{The overlapping region of the Bo\"otes field in $nW/m^2$ as measured in {\it Spitzer} at
%    3.6$\mu m$ (above) and {\it Herschel} at 250$\mu m$ (below)}
%    \label{fig:fullmaps}
%\end{center}
%\end{figure*}   

\subsection{Detected Source Masking}
\label{sec:masking}

We remove the detected sources from our maps so the cross-correlation is aimed at the unresolved fluctuations
in both sets of data.  We first generate  two masks, one for the {\it Spitzer} data and one for the {\it
Herschel} data (in all three bands). We combine these two masks to a common mask. This guarantees that both sets of data are
free of as much foreground contamination as possible, minimizing spurious correlations by having a
masked area in one map that is unmasked in the other. With a common mask we are also able to better handle
for mode-coupling effects which masking introduces. The final combined mask (a zoomed in portion can be
seen in Fig.~\ref{fig:mask}) removes 61.4\% of the pixels.

\subsubsection{{\it Spitzer}}

Since our model is based on sources below the detection threshold, we must pay particular attention
to generating a relatively deep mask for {\it Spitzer}. Otherwise unmasked {\it Spitzer} sources 
will correlate with the faint {\it Herschel} sources leading to a cross-correlation; in fact, as we find later,
a significant fraction of the cross-correlation is due to faint {\it Spitzer} sources correlating with
{\it Herschel} sources that are responsible for the SPIRE confusion noise.

We produce the {\it Spitzer} mask using a combination of catalogs from
the NOAO Deep Wide Field Survey (NDWFS) in the B$_w$, R and I bands, and {\sc SExtractor}) catalog 
that was generated for {\it Spitzer} maps at a
3-sigma detection threshold. As detailed in our previous work \citep{Cooray2012}, we start by iteratively running {\sc SExtractor}
with the same parameters used to generate the SDWFS catalogs \citep{Ashby2009}. This catalog is then
combined with the NDWFS catalogs in the bands listed above. Like with the {\it Herschel} mask, we
apply a flux cut, remove all pixels 5-sigma from the mean, and convolve everything with the PSF.
Finally, we combine this mask with the {\it Herschel} mask.

\begin{figure}
    \begin{center}
    \includegraphics[scale=0.48,clip]{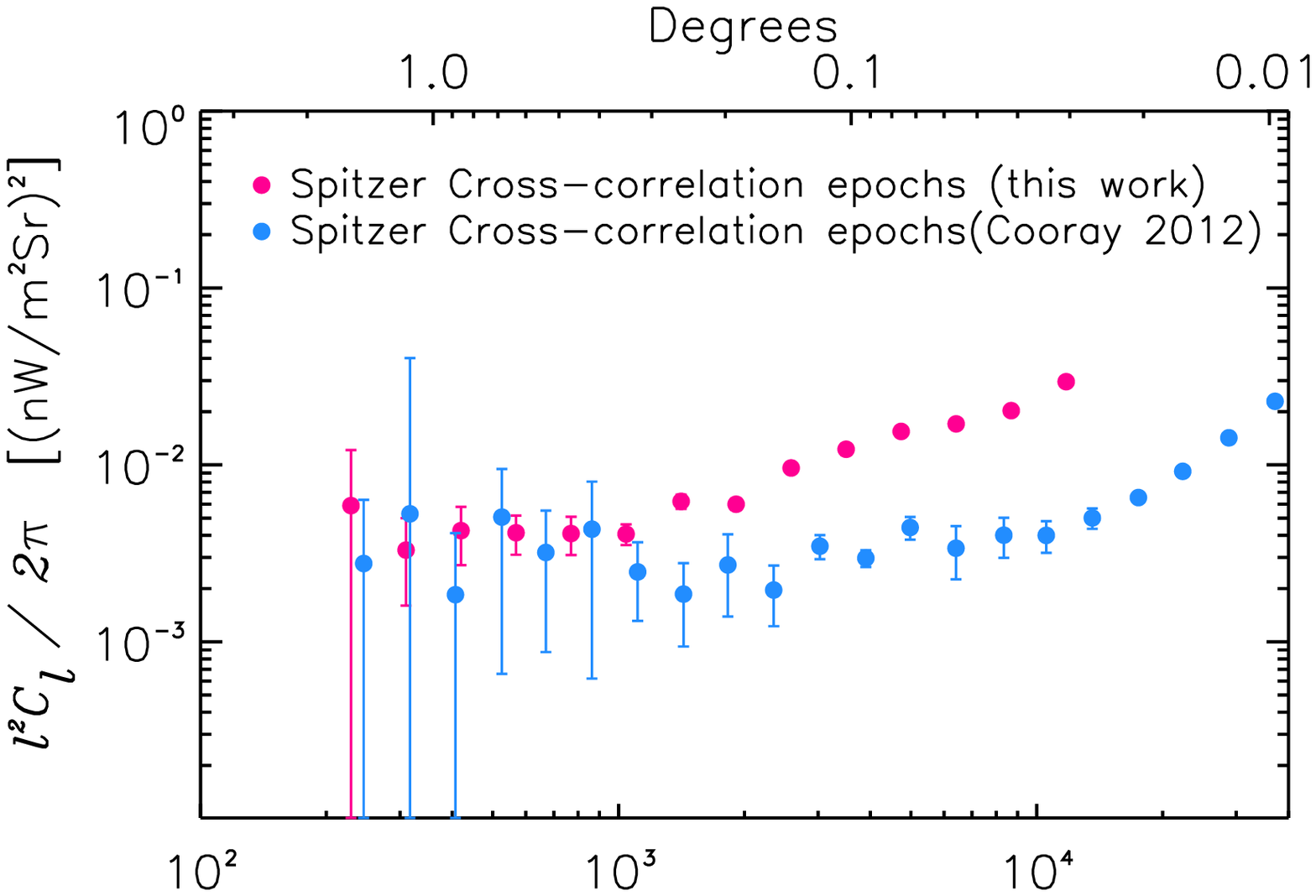}
    \caption[width=3in]{Comparison of {\it Spitzer} power spectra measured in the Bo\"otes field
    between the current and previous work \citep{Cooray2012}. In the previous work the power spectra
    measured is the cross-correlation of epochs at 1.2 arcsec/pixel. In this work, however, we have
    repixilized the maps to 6 arcsec/pixel and created a new mask, which we are not able to make as
    deep, or else we would end up with no data.}
\label{fig:spitzer_compare}
    \end{center}
\end{figure}

Since this image has been rebinned to a larger pixel size by a factor of five at 6 arcsec from our previous work
at 1.2 arcsec in \citet{Cooray2012}, significant blending occurs from neary galaxies that are resolved in the 1.2 arcsec
pixelized image. One effect of this is to increase the power at high-$\ell$ by increasing the shot-noise associated with unmasked galaxies.
An aggressive mask for {\it Spitzer}, consistent with \citet{Cooray2012}, cannot be used for this study since that will lead to
a small fraction of unmasked pixels for the cross-correlation study with {\it Herschel}. Our main limitation here is not
the depth of {\it Spitzer} imaging data, but the large PSF of {\it Herschel} maps. In  Fig.~\ref{fig:spitzer_compare} we compare the
power specrum from \citet{Cooray2012} and the new power spectrum of {\it Spitzer} imaging data with a mask that retains more of the faint
sources.

\subsubsection{Herschel} 
\label{herschelmask}

In the {\it Herschel} data, the mask is generated by a simple three step process in each of the SPIRE bands.
They are then combined into a single mask. The zeroth step is to crop the region in which {\it
Herschel} and {\it Spitzer} overlap into a rectangular area as shown in
Fig.~\ref{fig:crop_region}. First, we apply a flux cut at 50 mJy/beam, removing the
brightest galaxies and staying consistent with our previous work on the clustering of fluctuations in the SPIRE bands
\citep{Amblard:2011gc, Thacker2013}. Next, to include the extended nature of the sources, we expand
the mask by conlvolving with the point spread function (PSF).  Finally, we remove pixels with no
data, which arises mostly in 350 $\mu$m and 500 $\mu$m due to the enforced 6 arcsec pixel size
(oversampling the PSF), and pixels with corrupt data by applying a 5-sigma clip to the images.

\begin{figure}
    \begin{center}
    \includegraphics[scale=0.48,clip]{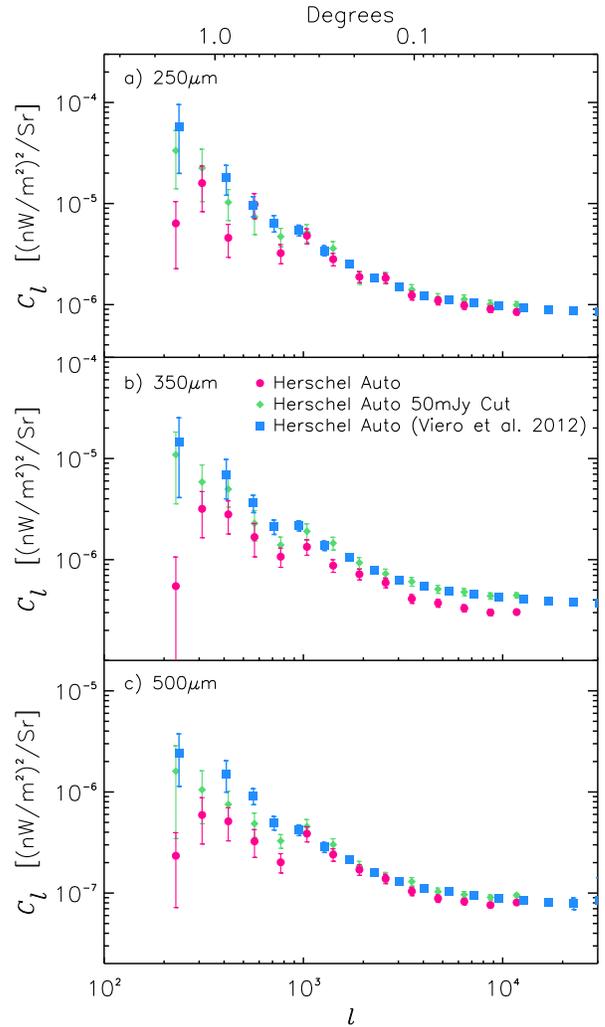}
    \caption[width=3in]{The comparison of {\it Herschel} auto-correlation power spectrum from our
        map (red points with a mask that removes 61.4\% of the pixels) to the measurement in
        \citet{Viero2013} (blue points). The latter is based on a source mask that removes SPIRE sources brighter
        than 50 mJy.  As discussed in Section~\ref{herschelmask} our effective flux density cut is
        at 29.5 mJy. In addition and just for comparison we also plot the auto-spectrum with a
    50mJy/Beam flux cut in green.}
\label{fig:her_comp}
    \end{center}
\end{figure}

The final mask for this study is obtained  by multiplying the individual
{\it Herschel} and {\it Spitzer} masks for the union between the two maps.
The auto power spectra we show in Fig.~\ref{fig:her_comp} use this combined mask between the two sets of data.
A comparison of the shot-noise levels at high-$\ell$
for the {\it Herschel} power spectrum with models for source counts reveals that our final mask has an effective depth in the
flux density cut at 250 $\mu$m of  29.5 mJy/beam. Fig.~\ref{fig:her_comp} shows the differences in the auto power spectra
with our combined mask vs.  power spectra measured with a mask based on a flux density cut for sources at 50 mJy.

\subsection{Angular Power Spectra and Sources of Error}
\label{sec:powerspec}

Our power spectra and cross power spectra measurements follow the procedures we have used for past
work \citep{Thacker2013, Cooray2012}.  Here we summarize the key ingredients related to the
uncertainties of the power spectrum measurements.

\subsubsection{Instrumental Noise}

One benefit of a cross-correlation is that instrumental noise is minimized,
especially between two different detectors or imaging experiments. The maps can be thought of as signal, $S$, plus a noise
component, $N$. So in a cross-correlation we have $M_1 \times M_2 = (S_1 + N_1) \times (S_2 + N_2)$.
Since the noise is expected to be random and because these are two different detectors, the noise should be
uncorrelated, thus dropping out of the cross-correlation. 

To get a handle on the instrumental noise component and an estimate of
systematic errors like zodiacal light or cirrus contamination, we perform jacknife tests. These tests
involve taking half maps for {\it Herschel} and single epoch maps for {\it Spitzer} and cross-correlating
their differences. For example in {\it Spitzer} we would take epochs $(1-2)$ cross correlated with epochs $(3-4)$, where
1 to 4 are the four epochs of the SWDFS survey. We average all combinations of this and find it is is stationary about zero. Finally, the variance arising
from this is an estimate of the residual noise and is added to the total error budget of the {\it Spitzer} auto power spectrum.

While it is assumed that the detector noise components of {\it Herschel} and {\it Spitzer} do not correlate, 
we still need to place a limit on any
residual noise correlations in the cross-correlation.
This is accomplished by taking null tests where we cross-correlate a {\it Herschel} map with all combinations of epoch differenced maps of
{\it Spitzer}. The variance of this multi epoch cross-correlations are  added to our error budget of the {\it Herschel} and {\it Spitzer} cross-correlation
power spectrum.

\subsubsection{Beam Correction}

At high $\ell$ there is a substantial drop in power due to the finite resolution of the detector and this drop in power
 needs to be accounted for. This is corrected by taking the power spectrum of the PSF (point
spread function) and normalizing to be one at low $\ell$. For cross spectra we need a correction
for the two separate beams. For that we  simply use the geometric mean of the two beams. For the {\it Herschel}
data we use the beam calculated by \citet{Amblard:2011gc} based on observations of
Neptune. The {\it Spitzer} beam used was calculated in \citet{Cooray2012} directly from
the PSF as described there.

\subsubsection{Mode-Coupling Correction}

\begin{figure}
    \begin{center}
    \includegraphics[scale=0.48,clip]{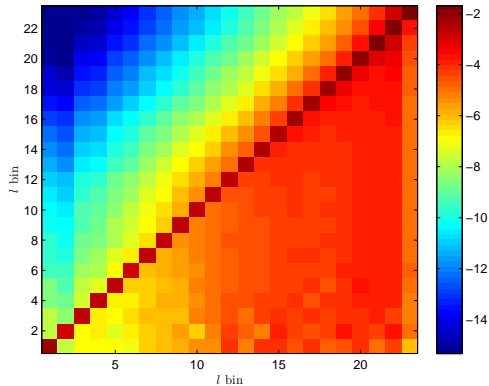}
    \caption[width=3in]{Mode-coupling matrix $M_{ll'}$ for the combined mask (log scale). }
\label{fig:mode}
    \end{center}
\end{figure}

Unlike the case of CMB, infrared background power spectra involves aggressive masks that remove a substantial fraction of the pixels.
A consequence of such masking is breaking up larger Fourier modes into smaller modes. This results in a shift of power
from low-$\ell$ to high-$\ell$. Using the method from \citet{Cooray2012} we generate a
mode-coupling matrix, $M_{\ell \ell'}$, shown in Fig.~\ref{fig:mode} by making maps of pure tones, masking them, and taking the
power spectrum. Thus, by construction the matrix transforms unmasked power to masked power and we
simply invert the matrix to obtain a transormation to an unbiased power spectrum. 

\subsubsection{Map Making Transfer Function}
\label{sec:transfer}

Unfortunately the map making process does not result in a perfect representation of the sky. The map
making process can induce ficticious correlations or suppress them based on deficiencies in the scan
pattern, the technique that was used to process the timeline data including any filtering that was
applied. We capture all of the modifications associated with the map making process by the transfer
function, $T(l)$. 

\begin{figure}
  \begin{center}
  \includegraphics[scale=0.48,clip]{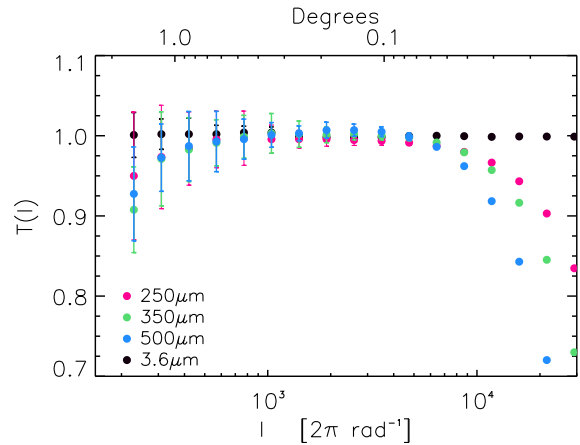}
  \caption[width=3in]{Map-making transfer function $T(l)$ for the {\it Herschel} and {\it Spitzer} Bo\"otes field
  anisotropy power spectrum. The uncertainties in the transfer function are calculated
  from 100 gaussian realizations of the sky as described in Section~\ref{sec:transfer}.
}
  \label{fig:transfer}
  \end{center}
\end{figure}

For {\it Herschel} data we calculate the transfer function by making 100 Gaussian random maps and
reading them into timeline data using the same pointing information as the actual data.  Then we
produce artificial maps from these timelines and take the average of the power spectra of each of
these maps to the power spectrum in each of the input maps.  For the {\it Spitzer} data, we follow a
similar approach but instead of reading data into timelines we create small tiles that are
re-mosaiced  using the same self-calibration algorithm as the actual data. These simulated tiles
include Gaussian fluctuations and instrumental noise consistent with the corresponding tile in the
actual data. Due to the dithering and tiling pattern that was used for actual observations in SDWFS
with {\it Spitzer}, the map-making transfer function is essentially one at all angular scales of
interest.  This is substantially different for {\it Herschel}/SPIRE with a transfer function that
departs from one at low and high $\ell$ as shown in Fig.~\ref{fig:transfer}. The difference at high
$\ell$ is due to the scan pattern that leaves stripes in the data; the difference at low $\ell$ or
large angular scales is due to the filtering that was applied to timeline data to remove gain drifts
during an individual scan.  That is the process used to make maps and to ensure the map-making
process has not generated a bias in the power spectrum.

\subsubsection{Power Spectrum Estimate}

To put these corrections to use, we must first define the cross-correlation. Given two maps H and S, 
the cross-correlation is calculated as:
\begin{equation}
    \langle C_{l_i} \rangle = \frac{\sum\limits_{l_1}^{l_2} w(l_x,l_y)
    \widetilde{H}(l_x,l_y) \widetilde{S}^*(l_x,l_y) } {\sum\limits_{l_1}^{l_2}w(l_x,l_y)},
\end{equation} 
where $\widetilde{H}$ and $\widetilde{S}$ are the 2D Fourier transforms of their respective maps and
$w(l_x,l_y)$ is the mask in Fourier space. The auto-spectra follows from the case when $H = S$. 

Finally, the final cross power spectrum estimate is obtained from the measured power spectrum, $\widetilde{C_{\ell'}}$,
with the above described corrections as:
\beq
C_\ell = \frac{M^{-1}_{\ell \ell'}\widetilde{C_{\ell'}} }{T_{\ell'}b_{\ell'}^{S} b_{\ell'}^{H}  } \, .
\eeq

To generate the Spitzer auto-spectrum (Fig \ref{fig:spitzer_compare}) we take the average of the
permuations of different cross-correlations involving the four epochs of SDWFS. For example, we calculate
$(1+2)\times(3+4)$ averaged with all other permuatations where 1 to 4 are the four epochs.
When compared to our previous work Spitzer auto-spectrum  deviates at high $\ell$. The higher
shot-noise is because we have repixelized the image from 1.2 arcsec/pixel in our prior work to 6
arcsec/pixel here. This also leads to a more shallow mask than our previous work.

Figure \ref{fig:PS} shows the cross-correlations with the best-fit model  (described below) shown for comparison. Again, we make use of
the four different {\it Spitzer} epochs in the cross-correlation. The cross-correlation is calculated as the average of 
(epoch$_i$ $+$ epoch$_j$)/2 $\times$ {\it Herschel} where $i \ne j$. 

In addition to the instrumental noise term and errors coming from the uncertainities in the beam function
and map-making transfer function, we also account for the cosmic variance in our total error budget.
More details are available in the  Appendix \ref{sec:cross_corr}.

\section{Halo Model}
 \label{sec:halomodel}

In this Section, we outline our model for the cross-correlation. The underlying model involves
three key components: far-IR galaxies, intra-halo light, and near-IR galaxies.

\subsection{Model for FIR background fluctuations from star-forming dusty galaxies}

We make use of the conditional luminosity function (CLF) models to calculate the power spectrum of FIR
galaxies \citep{Lee2009, DeBernardis2012}. The probability density of a halo/subhalo with mass $M$
to host a galaxy observed in a FIR band with luminosity L (i.e. $L_{250}$, $L_{350}$ and $L_{500}$ in this work) is given by
\be \label{eq:PLM}
P(L|M) = \frac{1}{\sqrt{2\pi}{\rm ln}(10)L\Sigma}{\rm exp}\left\{ -\frac{{\rm log_{10}}[L/\bar{L}(M,z)]^2}{2\Sigma^2}\right\},
\ee
where $\Sigma=0.3$ \cite{Lee2009} is the variance of ${\rm log_{10}}\bar{L}(M,z)$, and $\bar{L}(M,z)$ is the mean luminosity given a halo/subhalo mass $M$ at redshift $z$ which takes the form
\be \label{eq:LMz}
\bar{L}(M,z)=\bar{L}(M)(1+z)^p F_{\nu}[\nu_{\rm obs}(1+z)],
\ee
where we have
\be
\bar{L}(M)=L_0\left( \frac{M}{M_0}\right)^{\alpha} {\rm exp}\left[ -\left( \frac{M}{M_0}\right)^{\beta}\right].
\ee
Here we take $M_0=10^{12}$ $M_{\sun}$ consistent with previous work \citep{Lee2009,Cooray2012}, $L_0$, $\alpha$ and $\beta$ as free parameters to be
fitted by the data \footnote{We set $\beta=0$ when we perform the fitting process since we find
$\beta$ is close to zero as a free parameter \citep{DeBernardis2012}.}. We also consider the
redshift evolution of $\bar{L}(M)$ with a factor $(1+z)^p$, where $p$ is a free parameter. The
$F_{\nu}$ is spectral energy distribution (SED) for FIR galaxies, which takes into account the fact that the observed frequency $\nu_{\rm obs}$ comes from the frequency $\nu=\nu_{\rm obs}(1+z)$ at z. The SED can be expressed in terms of a modified blackbody normalized at 250 $\rm \mu m$
\be \label{eq:Fv}
F_{\nu} = \frac{(1-e^{-\tau})B(\nu,T_d)}{(1-e^{-\tau_0})B(\nu_0,T_d)},
\ee
where $B(\nu,T_d)$ is the blackbody spectrum, $T_d$ is the dust temperature of FIR galaxies, $\nu_0$ is the frequency of 250 $\rm \mu m$, $\tau=(\nu/\nu_0)^{\beta_d}$, $\tau_0=\tau(\nu_0)$ and $\beta_d$ is the dust emissivity spectral index. We fix $T_d=35\ \rm K$ and $\beta_d=2$ in this work which are consistent with results from observations \citep{Shang2011,Hwang2010,Chapman2006,Dunne2000,Amblard2010}.

Then the luminosity function (LF) can be obtained by
\be
\Phi(L,z)dL=dL\int dM P(L|M) n(M,z),
\ee
where $n(M,z)=n_h(M,z)+n_s(M,z)$ is the total halo mass function, and $n_h$ and $n_s$ are halo and
subhalo mass function respectively. In this work, we find that subhalos are not important and cannot
affect our results, so we ignore all subhalo terms in our analysis for simplicity, and we have
$n(M,z)\simeq n_h(M,z)$. Next, we can estimate the mean comoving emissivity of FIR galaxies,
$\bar{j}_{\nu}(z)$, at frequency $\nu$ and redshift $z$
\be \label{eq:jv}
\bar{j}_{\nu}(z) = \frac{1}{4\pi}\int \Phi(L,z) L dL.
\ee

Also, we can construct the halo occupation distribution (HOD) with the probability density shown by equation (\ref{eq:PLM}). Given a luminosity limit $L_{\rm min}$ determined by a survey,the average number of central galaxies hosted by halos of mass $M$ is
\be
\langle N_c(M)\rangle_{L\ge L_{\rm min}} = \int_{L_{\rm min}} P(L|M) dL.
\ee
The 1-halo and 2-halo terms of 3-D power spectrum for FIR galaxies are given by
\ba
P^{\rm 1h}_{\rm gg}(k,z) &=& \int dM n(M,z) \frac{\langle N_{\rm g}(N_{\rm g}-1)\rangle}{\bar{n}^2_{\rm g}}u^p(k|M,z),\label{eq:P1h} \\
P^{\rm 2h}_{\rm gg}(k,z) &=& \left[ \int dM b(M,z) n(M,z) \frac{\langle N_{\rm g}\rangle}{\bar{n}_{\rm g}} u(k|M,z) \right]^2\label{eq:P2h} \nonumber \\
                             & &\times P_{\rm lin}(k,z),
\ea
where $u(k|M,z)$ is the Fourier transform of the Navarro-Frenk-White (NFW) halo density profile for halos of mass $M$ at redshift $z$ \citep{Navarro1997}, and the power index $p$ takes $p=1$ when $\langle N_{\rm g}(N_{\rm g}-1)\rangle \le 1$ and $p=2$ otherwise \citep{Cooray2002}. $b(M,z)$ is the halo bias \citep{Sheth1999}, and $P_{\rm lin}$ is the linear matter power spectrum. The $\bar{n}_{\rm g}$ is the galaxy mean number density which is expressed as
\be
\bar{n}_{\rm g}(z) = \int dM n(M,z) \langle N_{\rm g}(M)\rangle.
\ee
Here $\langle N_{\rm g}(M)\rangle$ is the mean number of galaxies hosted by a halo of mass $M$.
Since we ignore the subhalo term in this work, we assume $\langle N_{\rm g}(M)\rangle\simeq \langle
N_c(M)\rangle$ and $\langle N_{\rm g}(N_{\rm g}-1)\rangle\simeq \langle N_c(M)\rangle^2$, which is a
good approximation in our calculation. 

The 2-D angular cross-power spectrum of FIR galaxies at observed frequencies $\nu$ and $\nu'$ can be obtained with the help of Limber approximation as
\be \label{eq:Cl}
C_{\ell,\rm FIR}^{\nu \nu'} = \int dz \left(\frac{d\chi}{dz}\right)\left( \frac{a}{\chi}\right)^2\bar{j}_{\nu}(z)\bar{j}_{\nu'}(z)P_{\rm gg}(k,z),
\ee
where $\chi$ is the comoving distance, $a$ is the scale factor, $\bar{j}_{\nu}(z)$ is the mean emissivity shown in equation (\ref{eq:jv}), and $P_{\rm gg}(k,z)=P_{\rm gg}^{\rm 1h}(k,z)+P_{\rm gg}^{\rm 2h}(k,z)$ is the galaxy power spectrum where $k=\ell/\chi$.

\subsection{Intra-Halo Light}

According to \citet{Cooray2012}, intra-halo light (IHL) mean luminosity for halos with mass $M$ at redshift $z$ is assumed to be
\be \label{eq:L_IHL}
\bar{L}_{\rm IHL}(M,z) = f_{\rm IHL}(M)L^{2.2}(M)(1+z)^{p_{\rm IHL}}F_{\lambda}^{\rm IHL},
\ee
where $f_{\rm IHL}(M)$ is the IHL fraction of the total halo luminosity, which has the form 
\be 
f_{\rm IHL}(M) = A_{\rm IHL}\left(\frac{M}{M_0}\right)^{\alpha_{\rm IHL}}.
\ee 
Here $A_{\rm IHL}$ is an amplitude factor and $\alpha_{\rm IHL}$ is a mass power index fixed to be
0.1 \citep{Cooray2012}. $L^{2.2}(M)=L^{2.2}_0(M)/\lambda_0$ is the total halo luminosity at 2.2
$\mu$m, where $\lambda_0=2.2\ \rm \mu m$ and $L^{2.2}_0(M)$ is given by \citep{Lin2004} 
\be
L^{2.2}_0(M) = 5.64\times10^{12}h_{70}^{-2}\left( \frac{M}{2.7\times10^{14}h_{70}^{-1}M_{\sun}}\right)^{0.72}\ L_{\sun}.
\ee
Then we can scale the total luminosity at $2.2\ \rm \mu m$ to the other wavelengths by the IHL SED $F_{\lambda}^{\rm IHL}$. Here the IHL SED is assumed to be the SED of old elliptical galaxies which are composed of old and red stars \citep{Krick2007}, and we normalize $F_{\lambda}^{\rm IHL}=1$ at $2.2\ \rm \mu m$.

The 1-halo and 2-halo terms of the 2-D IHL angular cross-power spectrum at
observed wavelengths $\lambda$ and $\lambda'$ can be estimated by
\ba
C_{\ell,\rm IHL}^{\lambda \lambda',\rm 1h} &=&\frac{1}{(4\pi)^2}\int dz \left( \frac{d\chi}{dz}\right) \left( \frac{a}{\chi}\right)^2 \nonumber \\
                                   &&\times \int dM n(M,z)u^2(k|M,z)\bar{L}_{\rm
                                   IHL}^{\lambda}\bar{L}_{\rm IHL}^{\lambda'},\label{eq:Cl1hIHL}\\
C_{\ell,\rm IHL}^{\lambda \lambda',\rm 2h} &=&\frac{1}{(4\pi)^2} \int dz \left( \frac{d\chi}{dz}\right) \left( \frac{a}{\chi}\right)^2 P_{\rm lin}(k,z)\nonumber \\
                                   &&\times \int dM b(M,z)n(M,z)u(k|M,z)\bar{L}_{\rm IHL}^{\lambda} \nonumber \\
                                   &&\times \int dM b(M,z)n(M,z)u(k|M,z)\bar{L}_{\rm
                                   IHL}^{\lambda'}.\label{eq:Cl2hIHL}
\ea

The total 2-D IHL angular cross-power spectrum is then given by
\be
C_{\ell,\rm IHL}^{\lambda \lambda'} = C_{\ell,\rm IHL}^{\lambda \lambda',\rm 1h} + C_{\ell,\rm IHL}^{\lambda \lambda',\rm 2h}.
\ee

\subsection{Model for NIR background fluctuations from known galaxy populations}

We follow \cite{Helgason12} to estimate the NIR background fluctuations from
known galaxy populations. We make use of their empirical fitting formulae of
luminosity functions (LFs) for measured galaxies in UV, optical and NIR bands
out to $z\sim 5$. Then, we estimate the mean emissivity $\bar{j}_{\nu}(z)$ at
the rest-frame frequencies and redshift to the observed frequency. We adopt the HOD model to calculate the 3-D galaxy power spectrum using equation (\ref{eq:P1h}) and (\ref{eq:P2h}) with $\langle N_{\rm g}\rangle=\langle N_c\rangle+\langle N_s\rangle$ and $\langle N_{\rm g}(N_{\rm g}-1)\rangle\simeq 2\langle N_s\rangle \langle N_c\rangle+\langle N_s\rangle^2$, where
\be
\langle N_c\rangle = \frac{1}{2}\left[ 1+{\rm erf}\left( \frac{{\rm log_{10}}M-{\rm log_{10}}M_{\rm min}}{\sigma_M}\right) \right],
\ee
and
\be
\langle N_s\rangle = \frac{1}{2}\left[ 1+{\rm erf}\left( \frac{{\rm log_{10}}M-{\rm log_{10}}2M_{\rm min}}{\sigma_M}\right) \right]\left( \frac{M}{M_s}\right)^{\alpha_s}.
\ee
Here $M_{\rm min}$ denotes the mass that a halo has $50\%$ probability of hosting a central galaxy,
and $\sigma_M$ is the transition width. We assume the satellite term has a cutoff mass with twice
the mass of central galaxy and grows as a power law with slope $\alpha_s$ and normalized by $M_s$.
We set $M_{\rm min}=10^9\ M_{\sun}$, $\sigma_M=0.2$, $M_s=5\times10^{10}\ M_{\sun}$, and
$\alpha_s=1$ \citep{Helgason12}. The 2-D angular cross-power spectrum $C_{\ell,\rm NIR}^{\nu\nu'}$
can be calculated by equation (\ref{eq:Cl}) with $\bar{j}_{\nu}(z)$ and $P_{\rm gg}(k,z)$ of
unresolved galaxies.

\begin{figure}
    \begin{center}
    \includegraphics[scale=0.48,clip]{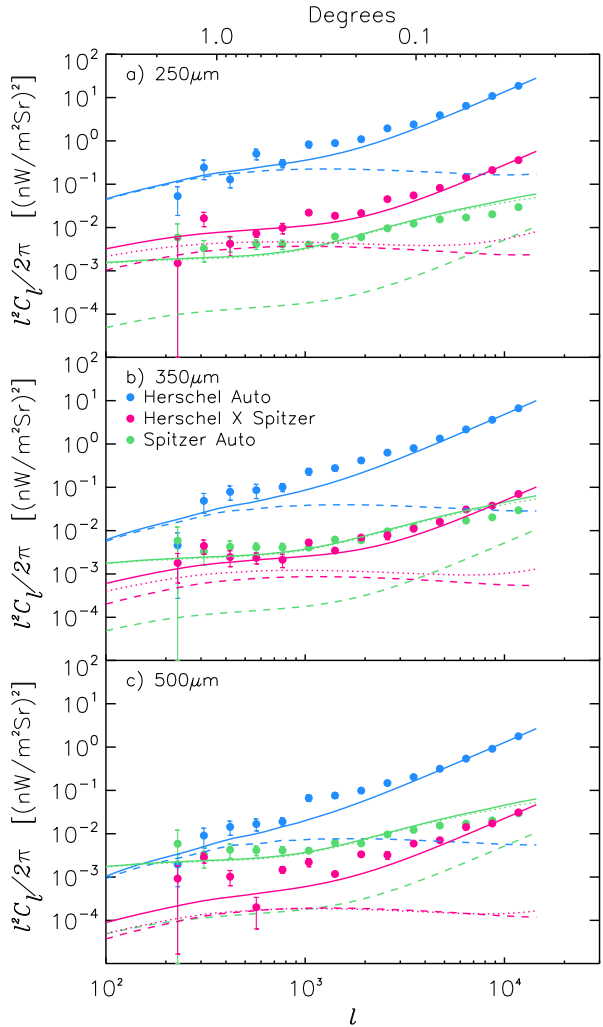}
    \caption[width=3in]{The auto and cross angular power spectra of Herschel (at $250$, $350$ and
    $500$ $\mu$m) and {\it Spitzer} (at 3.6 $\mu$m) surveys. The blue solid and dashed lines are the
    total fitting results and FIR galaxy power spectra, respectively. The green solid, dotted, and
    dashed are the fittings for the total, IHL and NIR galaxy power spectra. The pink solid line is the
    total cross-power spectrum fit while the dotted and dashed correspond to far-IR cross IHL and
    far-IR cross near-IR respectively.}
\label{fig:PS}
    \end{center}
\end{figure}
 
\begin{figure}
    \begin{center}
    \includegraphics[scale=0.45,clip]{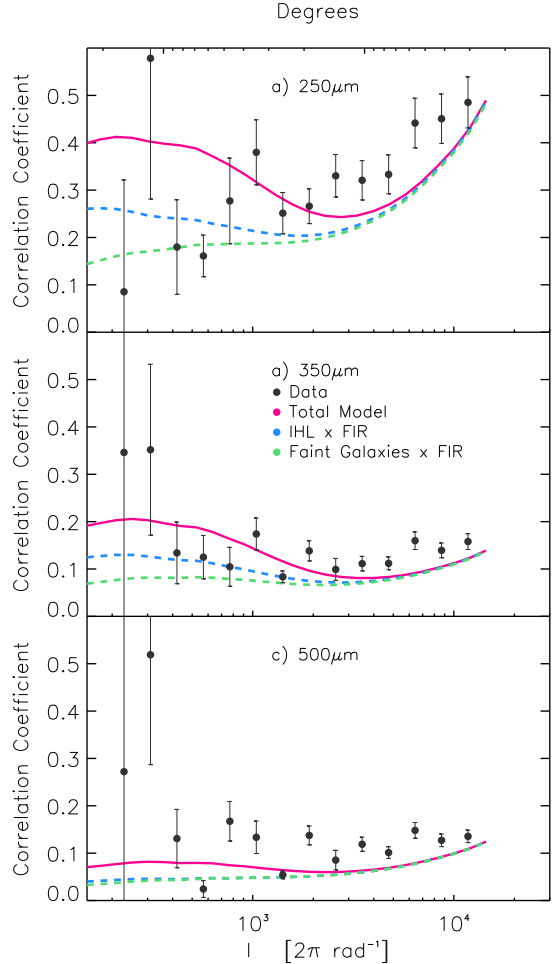}
    \caption[width=3in]{Correlation coefficient for the cross-correlation of Herschel and {\it Spitzer}
    (250, 350, 500 $\mu$m $\times$ 3.6 $\mu$m). The pink line corresponds to the total best fit halo
    model. The blue and green dashed lines show the component breakdown of the total halo model into
    the two terms that make up the total cross correlation. The blue line results from the
    correlation between IHL and FIR while the green line is from faint galaxies in the NIR
    correlated with FIR. } 
\label{fig:corr_coeff}
    \end{center}
\end{figure}    

\subsection{Model Comparison}

In this work, we have three angular auto power spectra $C_{\ell}^{\rm Her}$ in the far-infrared from
{\it Herschel}/SPIRE  at $250$, $350$ and $500\ \rm \mu m$, one angular auto power spectrum
$C_{\ell}^{\rm Spi}$ from {\it Spitzer} at $3.6\ \rm \mu m$, and three cross-power spectra
$C_{\ell}^{\rm cross}$ between {\it Herschel} and {\it Spitzer} data. Using the models discussed in the previous
Sections, we fit all these auto and cross power spectra with a single model.

As mentioned, we consider three components in our model to fit each dataset. For $C_{\ell}^{\rm
Her}$, we have $C_{\ell}^{\rm Her}=C_{\ell}^{\rm FIR}+C_{\ell,\rm shot}^{\rm
Her}$, where $C_{\ell}^{\rm FIR}=C_{\ell,\rm FIR}^{\rm \nu\nu'}$, $\nu_{\rm
obs},\nu'_{\rm obs}$ are the observed frequencies at $250$, $350$ or $500\ \rm \mu m$, and
$C_{\ell,\rm shot}^{\rm Her}$ is the shot-noise term. In a similar way, for $C_{\ell}^{\rm Spi}$,
we take $C_{\ell}^{\rm Spi}=C_{\ell}^{\rm NIR}+C_{\ell}^{\rm IHL}+C_{\ell,\rm shot}^{\rm Spi}$ at
$3.6\ \rm \mu m$. For $C_{\ell}^{\rm cross}$, it is expressed as $C_{\ell}^{\rm
cross}=C_{\ell}^{\rm FIR\times NIR}+C_{\ell}^{\rm FIR\times IHL}$. The model details of these cross powerspectra are outlined
in the Appendix. 

Since the shot-noise is a constant and dominant at high $\ell$, we derive the shot-noise terms
from the data at the high $\ell$ and fix them in the fitting process. The values of the shot-noise
terms we find are $C_{\ell,\rm shot}^{\rm Her}=8.4\times 10^{-7}, 3.0\times 10^{-7}$ and
$8.0\times 10^{-8}\ {\rm nW^2 m^{-4} sr^{-1}}$, and $C_{\ell,\rm shot}^{\rm cross}=1.7\times
10^{-8}, 3.0\times 10^{-9}$ and $1.4\times 10^{-9}\ {\rm nW^2 m^{-4} sr^{-1}}$ at $250$, $350$ and
$500\ \rm \mu m$, respectively. For $C_{\ell,\rm shot}^{\rm Spi}$, we scale the shot-noise of NIR
background $C_{\ell,\rm shot}^{\rm NIR}$ given by equation (13) in \cite{Helgason12} to match the
high-$\ell$ data of {\it Spitzer} at $3.6\ \rm \mu m$. This is done by adjust the minimum apparent
magnitude $m_{\rm min}$, and we find $m_{\rm min}=23$ which gives $C_{\ell,\rm shot}^{\rm
NIR}=3.1\times 10^{-10}\ {\rm nW^2 m^{-4} sr^{-1}}$.

We employ the Markov Chain Monte Carlo (MCMC) method to perform the fitting process. The
Metropolis-Hastings algorithm is adopted to determine the probability of accepting a new MCMC chain
point \citep{Metropolis1953}. We use $\chi^2$ distribution to calculate the likelihood function
$\mathcal{L}\propto {\rm exp}(-\chi_{\rm tot}^2/2)$. For the three datasets, we have $\chi^2_{\rm tot}=\chi^2_{\rm Her}+\chi^2_{\rm Spi}+\chi^2_{\rm cross}$, and the $\chi^2$ is given by
\be
\chi^2 = \sum^{N_{\rm d}}_{i=1} \frac{(C_{\ell}^{\rm obs}-C_{\ell}^{\rm th})^2}{\sigma_{\ell}^2},
\ee
where $N_{\rm d}$ is the number of the data points, $C_{\ell}^{\rm obs}$ and $C_{\ell}^{\rm th}$ are the angular power spectra from observation and theory, and $\sigma_{\ell}$ is the error for each data point.

For simplicity we take $z_{\rm min} = 0$, $z_{\rm max} = 6$,  $M_{\rm min}=10^{9}\ h^{-1}M_{\sun}$ and $M_{\rm max}=10^{14}\ h^{-1}M_{\sun}$ when we calculate the integral over 
redshift and halo mass in the power spectra. 
We use a uniform prior probability distribution for the free parameters in our model. The parameters
and their ranges are as follow: ${\rm log_{10}}L_0\in(-9,1)$, $\alpha\in(-5,5)$, $p\in(0,5)$ for the
FIR galaxy model, and $A_{\rm IHL}\in(-5,0)$, $p_{\rm IHL}\in(-5,5)$ for the IHL model. We generate
twelve parallel MCMC chains for each dataset, and collect about $120,000$ chain points after the
chains reach convergence. After the burn-in process and thinning the chains, we merge all the chains
together and get about $10,000$ chain points to illustrate the probability distribution of the free
parameters. The details of our MCMC method can be found in \cite{Gong2007}.

\section{Results and Discussion}

In this Section, we discuss the results from the data analysis and the MCMC model fits to the auto and cross power spectrum data. 
We will then outline estimates of derived  quantities from the model fitting results, such as the 
redshift distribution of the far and near-IR intensity and the cosmic dust density.

\subsection{Power spectra}

\begin{table}[!t]
\caption{The best-fit values and 1$\sigma$ errors of the model parameters from the MCMC constraints.}
\vspace{-4mm}
\begin{center}
\begin{tabular}{l | c | c | c  }
\hline\hline
           & 250 $\mu$m & 350 $\mu$m & 500 $\mu$m\\
\hline 
${\rm log_{10}}L_0$ & $-6.5\pm0.5$ & $-7.0\pm0.5$ & $-7.9\pm1.1$ \\ 
$\alpha$ & $0.23\pm0.06$ & $0.24\pm0.08$ & $0.22\pm0.11$ \\
$p$ & $3.9\pm0.12$ & $3.9\pm0.20$ & $3.9\pm0.25$\\
$\log_{10}A_{\rm IHL}$ & $-1.70\pm0.04$ & $-1.75\pm0.04$ & $-1.74\pm0.04$\\
$p_{\rm IHL}$ & $-3.2\pm0.4$ & $-2.5\pm0.3$ & $-2.6\pm0.3$\\
\hline
\end{tabular}
\end{center}
\label{tab:PS_fit}
\end{table}

In Figure \ref{fig:PS}, we show the angular auto and cross-power spectra of {\it Herschel} and {\it Spitzer}
in the Bo\"otes field. The top, middle and bottom panels are for {\it Herschel} 250, 350 and 500 $\mu$m, respectively.
The blue solid curve is total best-fit power spectrum for Herschel power spectrum, which is comprised
of two components, i.e. FIR galaxies (blue dashed), and total shot-noise
(not shown). Similarly, the green solid line is the best-fit for {\it Spitzer}
auto power spectrum, which has contributions given by IHL (green dotted) and NIR faint galaxies (green
dashed). As described in Cooray et al. (2012) the shot-noise of {\it Spitzer} power spectrum can be described by the faint galaxies, though
the clustering of such galaxies fall short of the fluctuations power spectrum at tens of arcminute angular scales.
The red solid line is the total cross correlation, including a shot-noise term not shown here, while the red dotted and dashed lines
separate the cross-correlation to the main terms given by IHL correlating with faint far-IR dusty galaxies and
faint near-IR galaxies correlating with faint far-IR dusty galaxies, respectively. In terms of the cross-correlations
we find  that these two terms are roughly comparable.

Another comparison of the data and model is shown in Fig. \ref{fig:corr_coeff}. Here, we calculate the
correlation coefficient separately from the data and compare to the correlation coefficient of the best-fit model. 
Displaying the information in this way is a valuable check of the relative strength of correlation. As the far-IR wavelength is increased from 250 $\mu$m to 500 $\mu$m
we find a less of a correlation between {\it Herschel} and {\it Spitzer}. Not only does the total correlation decrease, from our model fit we also find that the
 correlation between IHL and dusty far-IR galaxies, as a fraction of the total correlation, is also decreased.

The best-fit values and $1\sigma$ errors of the model parameters are shown in Table
\ref{tab:PS_fit}. We find the MCMC forces just a shot-noise term, i.e. a
straight line, to fit the data if we use data points out to $\ell\sim 10^5$ because the errors at
high $\ell$ (small scales) are very small compared to the errors at large scales. For these reasons, our results are obtained by fitting to
the power spectrum data at $\ell<10^4$, and ignoring all data points at $\ell>10^4$ in the fitting
process.

\subsection{Intensity redshift distribution}

\begin{figure}[t]
\includegraphics[trim=1cm 0 0 0,scale = 0.5,clip]{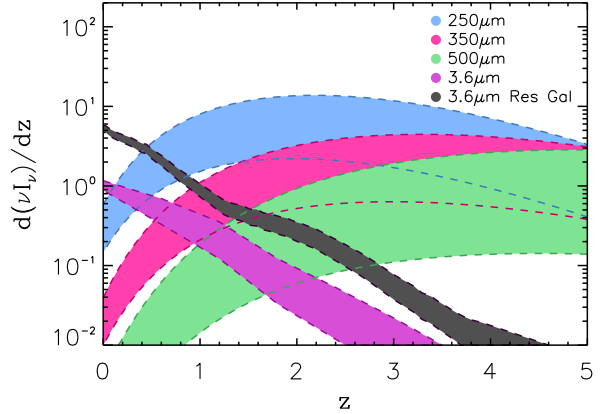}
\caption{\label{fig:dFdz} Redshift distribution of $d(\nu I_{\nu})/dz$ for Herschel FIR galaxies at
250, 350 and 500 $\mu$m, and {\it Spitzer} IHL and residual galaxies at 3.6 $\mu$m.}
\end{figure}

Using the fitting results of the best-fit model parameters and their errors, we  estimate the
redshift distribution of $d(\nu I_{\nu})/dz$ for far-IR dusty galaxies in the three SPIRE bands as
\be
\frac{d(\nu I_{\nu}^{\rm FIR})}{dz} = \frac{c}{H(z)(1+z)^2}\nu \bar{j}_{\nu}(z),
\ee  
where $c$ is the speed of light, $H(z)$ is the Hubble parameter, and $\bar{j}_{\nu}(z)$ is the mean emissivity given by equation (\ref{eq:jv}). Also, the redshift distribution of $d(\nu I_{\nu})/dz$ for the IHL component can be obtained by
\be
\frac{d(\nu I_{\nu}^{\rm IHL})}{dz} = \frac{c}{4\pi H(z)(1+z)^2} \int dM n(M,z) \bar{L}_{\rm IHL},
\ee
where $\bar{L}_{\rm IHL}(M,z)$ is the mean IHL luminosity given by equation (\ref{eq:L_IHL}), and $n(M,z)$ is the halo mass function. 

In Figure \ref{fig:dFdz}, we show $d(\nu I_{\nu})/dz$ as a function of the redshift for {\it Herschel} 
galaxies at 250, 350 and 500 $\mu$m, and {\it Spitzer} IHL and faint unresolved galaxies at 3.6 $\mu$m. We find the redshift
distribution has a turnover between $z=1$ and 2 for 250 $\mu$m, which indicates the intensity is
dominated by the sources at $z=1\sim2$ for 250 $\mu$m. For 350 $\mu$m, the turnover of $d(\nu
I_{\nu})/dz$ is not obvious and is shifted to higher redshift around $z=3$. For 500 $\mu$m, we find that the
sources at $z>3$  dominate the intensity. The increase in redshift with the wavelength for dusty
sources we find here is consistent with the well-known result in the literature
\citep{Bethermin2011,Amblard:2011gc,Viero2013}. On the other hand, the $d(\nu I_{\nu})/dz$ of {\it
Spitzer} IHL and residual galaxies at 3.6 $\mu$m has their maximum values at $z=0$ and decreases
quickly with increasing redshift. This shape, which is substantially different from {\it Herschel}
dusty galaxies, is the main reason that the cross-correlation signal between Herschel and {\it
Spitzer} is below 0.5 at 250 $\mu$m with a decrease to a value below 0.2 at 500 $\mu$m. The decrease
with increasing wavelength is consistent with the overall model description. If naively interpreted,
the small cross-correlation with  {\it Herschel}  could have been argued as evidence for a very
high-redshift origin for the {\it Spitzer} fluctuations, similar to the arguments that have been for
the origin of {\it Spitzer}-Chandra cross-correlation \citep{Cappelluti2012}. Our modeling suggest
the opposite: {\it Spitzer} fluctuations are very likely to be dominated by a source at $z < 1$
while intensity fluctuations in {\it Herschel} originate from $z=1$ and above at 250, 350 and 500
$\mu$m.

\subsection{Cosmic Dust Density}

\begin{figure}[t]
\includegraphics[scale = 0.52,trim=1.2cm 0 0 0,clip]{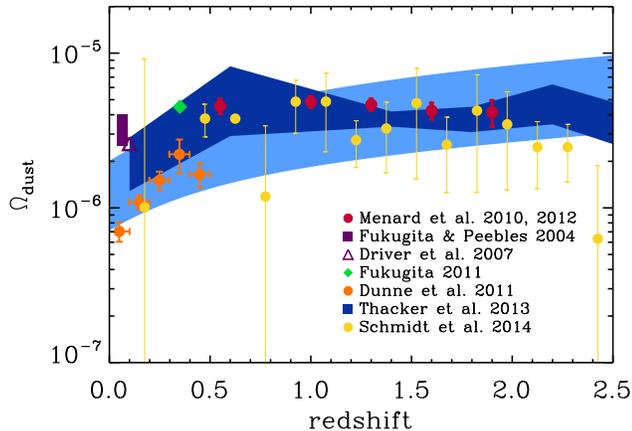}
\caption{\label{fig:Omega_dust} The fractional cosmic dust density $\Omega_{\rm
dust}$ as a function of redshift from FIR galaxies. The light blue shaded region shows the 1$\sigma$
range of $\Omega_{\rm dust}$ for FIR galaxies derived from our MCMC fitting results in this paper. We also show
the results from different observations for $\Omega_{\rm dust}$. The orange squares are the
measurement of H-ATLAS dust mass function in \cite{Dunne10}, and the other data are based on the
extinction measurements of the SDSS and 2dF surveys
\citep{Menard2010,Menard2012,Fukugita2004,Fukugita2011,Driver2007}. We also include recent
measurements from \citet{Schmidt2014} using the cross-correlation between the Planck High Frequency
Instrument and quasars from the Sloan Digital Sky Survey DR7. Finally, we include results from
\citet{Thacker2013} as the dark blue shaded region.}  
\end{figure}

As an application of our models we also derive the fractional cosmic dust density $\Omega_{\rm dust}$ of galaxies from the
fitting results. Following \cite{Thacker2013}, the $\Omega_{\rm dust}$ for galaxies is given by 
\be
\Omega_{\rm dust}(z) = \frac{1}{\rho_{0}}\int dL\Phi(L,z)M_{\rm dust}(L),
\ee
where $\rho_0$ is the current cosmic critical density, $\Phi(L,z)$ is the luminosity function, and
$M_{\rm dust}$ is the dust mass with IR luminosity $L$ with temperature assumed to be 35$K$ that we
fix in the power spectrum model. We make use of equation (4) of \cite{Fu2012} to estimate $M_{\rm
dust}$ from IR luminosity \citep{Thacker2013}. Also, we assume the dust opacity $\kappa_d$ takes the
form as $\kappa_d=A_{\kappa}\nu^{\beta_d}$, where $\beta_d=2$ is the dust emissivity spectral index,
and $A_{\kappa}$ is the normalization factor which is estimated by $\kappa_d=0.07\pm0.02\ \rm
m^2kg^{-1}$ at 850 $\mu$m \citep{Dunne2000,James2002}.  

We show $\Omega_{\rm dust}$ as a function of redshift in Figure
\ref{fig:Omega_dust}. The blue shaded region is the 1$\sigma$ range of the $\Omega_{\rm dust}$ from
the FIR. We also show the other measurements for comparison. The orange squares are from the
H-ATLAS dust mass function as measured in \cite{Dunne10}, and the other data are based on the extinction
measurements of the SDSS and 2dF surveys
\citep{Menard2010,Menard2012,Fukugita2004,Fukugita2011,Driver2007}. We find our $\Omega_{\rm dust}$
result from galaxies is consistent with the other measurements, which has $\Omega_{\rm
dust}$ increasing at higher redshift.

\section{Summary}

We have calculated the cross-correlation power spectrum of the cosmic infrared
background at far and near-IR wavelengths using {\it Herschel} and {\it Spitzer} in the Bo\"otes field. We measured the
correlation coefficient to be between 10-40\% with the highest correlation seen in the 250 $\mu$m
band and the lowest in 500 $\mu$m.

Recent results from \citet{Cooray2012} and \citet{Zemcov2014} suggest that the near-IR background
anisotropies have a mostly low redshift origin at $z < 1$, arising from intra-halo light and faint dwaft
galaxies. Meanwhile the far-IR signal is dominated by dusty galaxies peaking at a redshift of $\sim$1 and above
\citep{Amblard2010,Thacker2013, Viero2013}. By cross correlating {\it Herschel} with {\it Spitzer}
we are able to provide a check on the robustness of such a model. We find that not only can such a
model fit the auto-correlations they were designed to fit, but can also explain the
cross-correlation signal, including the wavelength dependence of the cross-correlation coefficient. 

\acknowledgements
This work was supported by NSF AST-1313319 and GAANN at UCI.  The public Herschel data in the
Herschel Science Archive was taken by HerMES with observation Id's 1342187711 to 1342187713,
1342188090, 1342188650 to 1342188651, 1342188681 to 1342188682, and 1342189108. HIPE was a joint
development by the Herschel Science Ground Segment Consortium, consisting of ESA, the NASA Herschel
Science Center, and the HIFI, PACS and SPIRE consortia. We thank Rick Arendt for sharing of his
self-calibration pipeline for Spitzer data.  We thank ESA for maintaining the Herschel Science
Archive and for making available reduced and calibrated time-ordered data of every SPIRE scan that
was done during the Herschel mission. 

\appendix
\section{Cross Power Spectra}

Here we discuss the 2-D angular cross-power spectra of Herschel and {\it Spitzer} in our model. The total cross-power spectrum $C_{\ell}^{\rm cross}$ is composed of four components which are 
\be
C_{\ell}^{\rm cross} = C_{\ell}^{\rm FIR\times NIR} + C_{\ell}^{\rm FIR\times IHL}.
\ee
Here $C_{\ell}=\nu_{\rm obs}\nu'_{\rm obs}C_{\ell}^{\nu\nu'}$ where $\nu_{\rm obs}$ is the observed frequency at $250$, $350$ or $500\ \rm \mu m$ of Herschel survey, and $\nu'_{\rm obs}$ is the observed frequency at $3.6\ \rm \mu m$ of {\it Spitzer} survey.

The 1-halo and 2-halo terms of $C_{\ell,\rm FIR\times NIR}^{\nu\nu'}$ are given by
\ba
C_{\ell,\rm \rm FIR\times NIR}^{\nu\nu', \rm 1h} &=& \int dz \left(\frac{d\chi}{dz}\right)\left( \frac{a}{\chi}\right)^2\bar{j}_{\nu}(z)\bar{j}_{\nu'}(z)\int dM n(M,z) \frac{\sqrt{\langle N_{\rm g}(N_{\rm g}-1)\rangle_{\nu}} \sqrt{\langle N_{\rm g}(N_{\rm g}-1)\rangle_{\nu'}}}{\bar{n}^{\nu}_{\rm g}(z)\bar{n}^{\nu'}_{\rm g}(z)}\sqrt{u^{p_{\nu}}}\sqrt{u^{p_{\nu'}}},\\
C_{\ell,\rm \rm FIR\times NIR}^{\nu\nu', \rm 2h} &=& \int dz \left(\frac{d\chi}{dz}\right)\left( \frac{a}{\chi}\right)^2\bar{j}_{\nu}(z)\bar{j}_{\nu'}(z)\left[\int dM b(M,z) n(M,z) \frac{\langle N_{\rm g}\rangle_{\nu}}{\bar{n}_{\rm g}^{\nu}(z)} u(k|M,z) \right] \nonumber \\
             && \times \left[\int dM b(M,z) n(M,z) \frac{\langle N_{\rm g}\rangle_{\nu'}}{\bar{n}_{\rm g}^{\nu'}(z)} u(k|M,z) \right]P_{\rm lin}(k,z).
\ea
Here $\nu$ and $\nu'$ denote FIR and NIR bands for Herschel and {\it Spitzer} respectively. Then we have $C_{\ell,\rm \rm FIR\times NIR}^{\nu\nu'}=C_{\ell,\rm \rm FIR\times NIR}^{\nu\nu', \rm 1h}+C_{\ell,\rm \rm FIR\times NIR}^{\nu\nu', \rm 2h}$. 

The 1-halo and 2-halo terms of $C_{\ell,\rm FIR\times IHL}^{\nu\nu'}$ are
\ba
C_{\ell,\rm \rm FIR\times IHL}^{\nu\nu', \rm 1h} &=& \frac{1}{4\pi}\int dz \left(\frac{d\chi}{dz}\right)\left( \frac{a}{\chi}\right)^2\bar{j}_{\nu}(z)\int dM n(M,z) \frac{\sqrt{\langle N_{\rm g}(N_{\rm g}-1)\rangle_{\nu}}}{\bar{n}^{\nu}_{\rm g}(z)}\sqrt{u^{p_{\nu}}(k|M,z)}\bar{L}_{\rm IHL}^{\nu'}(M,z)u(k|M,z),\label{eq:Cl1h_FIRxIHL}\\
C_{\ell,\rm \rm FIR\times IHL}^{\nu\nu', \rm 2h} &=& \frac{1}{4\pi}\int dz \left(\frac{d\chi}{dz}\right)\left( \frac{a}{\chi}\right)^2\bar{j}_{\nu}(z)\left[\int dM b(M,z) n(M,z) \frac{\langle N_{\rm g}\rangle_{\nu}}{\bar{n}_{\rm g}^{\nu}(z)} u(k|M,z) \right] \nonumber \\
                                   &&\times \left[\int dM b(M,z)n(M,z)u(k|M,z)\bar{L}_{\rm IHL}^{\nu'}(M,z)\right]P_{\rm lin}(k,z).\label{eq:Cl2h_FIRxIHL}
\ea
So we get $C_{\ell,\rm \rm FIR\times IHL}^{\nu\nu'}=C_{\ell,\rm \rm FIR\times IHL}^{\nu\nu', \rm 1h}+C_{\ell,\rm \rm FIR\times IHL}^{\nu\nu', \rm 2h}$. 

\section{Flat Sky Approximation} 
\label{sec:Flat}

Using the formalism of \citet{Hivon2002}, we show how to get the flat sky approximation
which allows the use of Fourier transforms instead of spherical harmonics to calculate the power
spectrum. The flat sky approximation requires $\theta \ll 1$ and $\ell \gg 1$. We start by defining
the weighted sum over the multipole moments as
\beq
M(\vec{\ell}) = \sqrt{\frac{4\pi}{2\ell+1}}\sum\limits_{m=-\ell}^{\ell} i^{-m}M_{\ell m}e^{i m
\phi_\ell}.
\eeq

For this derivation we need the following approximations for when $\theta \ll 1$ and $\ell \gg 1$.
The first is known as the Jacobi-Anger expansion of the plane wave,
\beq
e^{i\vec{\ell}\cdot\vec{r}} = \sum\limits_{m}i^mJ_m(\ell \theta) e^{i m (\phi - \phi_\ell)},
\eeq
where $J_m(\ell \theta)$ are Bessel functions and we use the fact that since $\theta$ is small r
$\approx \theta$.

Next, to show that $Y_{\ell m} $ is approximately $\sqrt{\ell/2\pi} J_m(\ell m) e^{i m
\phi}$, we start with the general Legendre equation and perform a change of variables $\cos(\theta)
\Rightarrow \cos(\frac{\theta'}{\ell})$. Using the fact that $\theta'/\ell \ll 1$ to simplify, and
multiply the equation by $\theta'^2 / \ell^2$ to get it into the correct form. Now, one can
recognize the equation as Bessel's equation with solution $J_m(\theta') = J_m(\ell \theta)$

Putting this all together, we decompose the maps into spherical harmonics and introduce some
simplifications to obtain,
\begin{align}
M(\hat{n}) &= \sum\limits_{m \ell} M_{\ell m} Y_{\ell m}, \\
&=  \sum\limits_{m \ell} M_{\ell m} \sqrt{\frac{\ell}{2\pi}} J_m(\ell m) e^{i m \phi}\nonumber, \\ 
&=  \sum\limits_{m \ell} \frac{\ell}{2\pi}\int \frac{\mathrm{d}\phi_{\ell}}{2\pi}M(\vec{\ell}) i^m
J_m(\ell m) e^{i m (\phi-\phi_{\ell})} \nonumber,\\
&\approx \int \frac{\mathrm{d}^2\ell}{(2\pi)^2} M(\vec{\ell}) e^{i\vec{\ell}\cdot\vec{r}} \nonumber.
\end{align} 

\section{Cross Correlation Power Spectrum and Cosmic Variance}
\label{sec:cross_corr}

Analogous to the convolution theorem, correlations obey a similar relation except with a complex
conjugate. Below, $\mathcal{F}$ denotes 2D Fourier transforms and $\star$ denotes cross-correlation,
where M$_i$ are 2D maps:  
\beq
\mathcal{F}(M_{1} \star M_{2}) =  \mathcal{F}(M_1) \cdot \mathcal{F}(M_2)^* \, .
\eeq

Since we can use Fourier transforms as a suitable basis to obtain a power spectrum (see
\ref{sec:Flat} in Appendix), we are guaranteed that the cross-correlation above is also a power
spectrum. Putting it together, for a specific $\ell_i$ bin and including Fourier masking, we obtain
\begin{equation}
    \langle C_{l_i} \rangle = \frac{\sum\limits_{l_1}^{l_2} w(l_x,l_y)
    \widetilde{M_1}(l_x,l_y) \widetilde{M_2}^*(l_x,l_y) } {\sum\limits_{l_1}^{l_2}w(l_x,l_y)}.
\end{equation}
Here $\widetilde{M_1} = \mathcal{F}(M_1)$ and $w(l_x,l_y)$ is a Fourier mask that
has the value one for modes we keep and zero for modes we mask. The binning is
done in annular rings so $l_1 \ge l_x^2+l_y^2$ and $l_2 \le l_x^2+l_y^2$.

Cosmic variance , $\delta C_\ell$, is the expected variance on the power spectrum estimate at each $\ell$ mode and is
given by
\begin{equation}
    \delta C_\ell = \sqrt{ \frac{2}{f_{\mathrm{sky}}(2\ell+1)\Delta\ell}} (C_\ell^{\mathrm{auto}} +
    N_\ell),
\end{equation}
where $N_\ell$ is the noise power spectrum generated through jacknife and similar
techniques, $f_{\mathrm{sky}}$ is the fraction of the sky covered by the map, and $\Delta\ell$ is
the width of the $\ell$-bin.

For the cross-correation spectrum  the cosmic variance is
\begin{equation}
    \delta C_\ell = \sqrt{ \frac{1}{f_{\mathrm{sky}}(2\ell+1)\Delta\ell} \left[
        (C_{A \ell}^{\mathrm{auto}} + N_{A \ell})(C_{B \ell}^{\mathrm{auto}} + N_{B \ell}) +
\left(C_\ell^{\mathrm{A}\times\mathrm{B}}\right)^2  \right]},
\end{equation}
where $C_\ell^{\mathrm{A}\times\mathrm{B}}$ is the cross-correlation power spectrum.

%\begin{table}
%\begin{center}
%\begin{tabular}{ |c | c| c|  }
%\hline
%  Wavelength & Atlas (Jy$^2$/Sr) \\
%\hline
%  250 & $6382 \pm 447$ \\
%  350 & $4381 \pm 339$ \\
%  500 & $1893 \pm 158$ \\
%\hline
%\end{tabular}
%\label{tab:Poisson}
%\caption[width=3in]{Best fit Poisson Noise level for each field from the halo model. The errors represent the 68\% confidence region.}
%\end{center}
%\end{table} 


\begin{thebibliography}{}

\bibitem[Amblard et al., 2010]{Amblard2010}
Amblard, A., Cooray, A., Serra, P., et al. 2010, A\&A, 518, L9


\bibitem[{Amblard} {et~al.}, 2011]{Amblard:2011gc}
{Amblard}, A., {et~al.} 2011, \nat, 470, 510

\bibitem[{{Arendt} {et~al.}(2000){Arendt}, {Fixsen}, \& {Moseley}}]{Arendt2000}
{Arendt}, R.~G., {Fixsen}, D.~J., \& {Moseley}, S.~H. 2000, \apj, 536, 500

\bibitem[{{Ashby} {et~al.}(2009){Ashby}, {Stern}, {Brodwin}, {Griffith},
  {Eisenhardt}, {Koz{\l}owski}, {Kochanek}, {Bock}, {Borys}, {Brand}, {Brown},
  {Cool}, {Cooray}, {Croft}, {Dey}, {Eisenstein}, {Gonzalez}, {Gorjian},
  {Grogin}, {Ivison}, {Jacob}, {Jannuzi}, {Mainzer}, {Moustakas},
  {R{\"o}ttgering}, {Seymour}, {Smith}, {Stanford}, {Stauffer}, {Sullivan},
  {van Breugel}, {Willner}, \& {Wright}}]{Ashby2009}
{Ashby}, M.~L.~N., {et~al.} 2009, \apj, 701, 428

\bibitem[{{Bertin} \& {Arnouts}(1996)}]{Bertin}
{Bertin}, E. \& {Arnouts}, S. 1996, \aaps, 117, 393


\bibitem[Bethermin et al., 2011]{Bethermin2011}
Bethermin, M., Dole, H., Lagache, G., et al. 2011, A\&A, 529, A4


\bibitem[{{Cantalupo} {et~al.}(2010){Cantalupo}, {Borrill}, {Jaffe}, {Kisner},
  \& {Stompor}}]{Cantalupo2010}
{Cantalupo}, C.~M., {et~al.} 2010, \apjs, 187, 212

\bibitem[Cappelluti et al., 2012]{Cappelluti2012}
Cappelluti, N., Kashlinsky, A., and Arendt, R.~G., et al. 2012, ApJ, 769, 68

\bibitem[Chapman \& Wardle, 2006]{Chapman2006}
Chapman, J.~F.~ \& Wardle, M.~ 2006 MNRAS, 371, 513

\bibitem[Carollo et al.(2010)]{Carollo10}
Carollo, D., et al. 2010, ApJ, 712, 692–727


\bibitem[Cooray \& Sheth, 2002]{Cooray2002}
 Cooray, A., Sheth, R.K. 2002, PR,  372, 1


\bibitem[{{Cooray} {et~al.}(2012){Cooray}, {Smidt}, {De Bernardis}, {Gong},
  {Stern}, {Ashby}, {Eisenhardt}, {Frazer}, {Gonzalez}, {Kochanek},
  {Kozlowski}, \& {Wright}}]{Cooray2012}
{Cooray}, A., {et~al.} 2012, Nature, 490, 514


\bibitem[Courteau et al.(2011)]{Courteau11}
Courteau, S., et al. 2011, ApJ, 739, 20


\bibitem[De Bernardis \& Cooray, 2012]{DeBernardis2012}
De Bernardis, F., \& Cooray, A.~ 2012, ApJ, 760, 14
                                
\bibitem[{{Dole} {et~al.}(2006){Dole}, {Lagache}, {Puget}, {Caputi},
  {Fern{\'a}ndez-Conde}, {Le Floc'h}, {Papovich}, {P{\'e}rez-Gonz{\'a}lez},
  {Rieke}, \& {Blaylock}}]{Dole2006}
{Dole}, H., {Lagache}, G., {Puget}, J.-L., et al. 2006, \aap, 451, 417 

\bibitem[Driver et al., 2007]{Driver2007}
Driver, S. P., et al. 2007, MNRAS, 379, 1022


\bibitem[Dunne et al., 2000]{Dunne2000}
Dunne, L.~, Eales, S.~A.~, Edmunds, M.~G.~, et al. 2000, MNRAS, 315, 115


\bibitem[Dunne et al., 2010]{Dunne10}
Dunne, L.~, Gomez, H.~, da Cunha, E.~ S., et al. 2010, MNRAS, 417, 1510

\bibitem[Fu et al.(2012)]{Fu2012}
Fu, H., Jullo, E., Cooray, A., et al. 2012, ApJ, 753, 12


\bibitem[Fukugita \& Peebles, 2004]{Fukugita2004}
Fukugita, M., Peebles P. J. E., 2004, ApJ, 616, 643

\bibitem[Fukugita et al., 2011]{Fukugita2011}
Fukugita, M., 2011, arXiv, arXiv:1103.4191


\bibitem[Gonzalez et al.(2005)]{Gonzalez05}
Gonzalez, A. H., Zabludoff, A. I. \& Zaritsky, D. 2005, ApJ, 618, 195-213

\bibitem[Gong \& Chen(2007)]{Gong2007}
Gong, Y., \& Chen, X. 2007, PRD, 76, 123007

\bibitem[Griffin et al., 2010]{Griffin:2010hp}
 Griffin, M.~J.~, Abergel, M.~J.~, Abreu, A.~, et al. 2010,  A\&A, 518, L3

\bibitem[{{Helgason} {et~al.}(2012){Helgason}, {Ricotti}, \&
  {Kashlinsky}}]{Helgason12}
{Helgason}, K., {Ricotti}, M., \& {Kashlinsky}, A. 2012, \apj, 752, 113

\bibitem[{{Hivon} {et~al.}(2002){Hivon}, {G{\'o}rski}, {Netterfield}, {Crill},
  {Prunet}, \& {Hansen}}]{Hivon2002}
{Hivon}, E., {et~al.} 2002, \apj, 567, 2


\bibitem[Hwang et al., 2010]{Hwang2010}
Hwang, H.S., Elbaz, D.~, Magdis, G.~, et al., 2010, MNRAS, 409, 75


\bibitem[James et al., 2002]{James2002}
James, A.~, Dunne, L. et al., 2002 MNRAS, 335, 753

\bibitem[Kashlinsky et~al. (2005)]{Kashlinsky2005}
Kashlinsky, A., Arendt, R.~G., Mather, J., \& Moseley, S.~H. 2005, Nature, 438, 45 

\bibitem[Kashlinsky et~al. (2007)]{Kashlinsky2007}
Kashlinsky, A., Arendt, R.~G., Mather, J., \& Moseley, S.~H. 2007, ApJL, 654, L5

\bibitem[Kashlinsky et~al. (2012)]{Kashlinsky2012}
Kashlinsky, A., Arendt, R.~G., Ashby, M.~L.~N., Fazio, G.~G., Mather, J., \& Moseley S.~H. 2012, ApJ, 753, 63 



\bibitem[Krick \& Bernstein(2007)]{Krick2007}
Krick, J. E., \& Bernstein, R. A. 2007, ApJ, 134, 466-493


\bibitem[Lee et al. (2009)]{Lee2009}
Lee {\it et al.}, ApJ 695, 368 (2009)


\bibitem[Lin et al., (2004)]{Lin2004}
Lin, Y. T., Mohr, J. J., \& Stanford, S. A. 2004, ApJ, 610, 745-761


\bibitem[Matsumoto et~al. (2011)]{Matsumoto2011}
Matsumoto, T., Seo, H.~J., Jeong, W.-S., Lee, H.~M., Matsuura, S., Matsuhara, H., Oyabu, S., Pyo, J., et al. 2011, ApJ, 742, 124


\bibitem[Menard et al., 2010]{Menard2010}
Menard B., Scranton R., Fukugita M., Richards G., 2010, MNRAS, 405, 1025


\bibitem[Menard \& Fukugita (2012)]{Menard2012}
Menard, B., \& Fukugita, M. 2012, ApJ, 754, 116


\bibitem[Metropolis et al.(1953)]{Metropolis1953}
Metropolis, N., Rosenbluth, A. W., Rosenbluth, M. N., Teller, A. H., \& Teller,
E. 1953, JCP, 21, 1087


\bibitem[Navarro et al., 1997]{Navarro1997}
Navarro, J. F., Frenk, C. S., White, S. D. M. 1997, ApJ, 490, 493

\bibitem[Oliver et~al. (2010)]{Oliver2010}
{Oliver}, S.~J., {Wang}, L., {Smith}, A.~J., {Altieri}, B., {Amblard}, A., et al. 2010, \aap, 518, L21 

\bibitem[{{Ott}(2010)}]{Ott2010}
{Ott}, S. 2010, in Astronomical Society of the Pacific Conference Series, Vol.
  434, Astronomical Data Analysis Software and Systems XIX, ed. Y.~{Mizumoto},
  K.-I. {Morita}, \& M.~{Ohishi}, 139

\bibitem[Pillbratt et al., 2010]{Pillbratt2010}
Pillbratt, G. L., et al. 2010, A\&A, 518, L1

\bibitem[{Planck Collaboration (2014)}]{Planck2014}
{Planck Collaboration} and {Ade}, P.~A.~R., {Aghanim}, N., {Armitage-Caplan}, C., {Arnaud}, M.,
{Ashdown}, M., {Atrio-Barandela}, F., {Aumont}, J., {Baccigalupi}, C., {Banday}, A.~J., et al.
2014,\aap,571, A16    
    

\bibitem[Purcell et al.(2008)]{Purcell08}
Purcell, C. W., Bullock, J. S. \& Zentner, A. R. 2008, MNRAS, 391, 550-558

\bibitem[Schmidt et al., 2014]{Schmidt2014}
    Schmidt, S., Menard, B., et al., 2014, MNRAS 446, 2696

\bibitem[Shang et al., 2011]{Shang2011}
Shang, C., et al. 2011, arXiv:1109.1522


\bibitem[Sheth \& Tormen, 1999]{Sheth1999}
Sheth, R. K., Tormen, G. 1999, MNRAS, 308, 119



\bibitem[{{Thacker} {et~al.}(2013){Thacker}, {Cooray}, {Smidt}, {De Bernardis},
  {Mitchell-Wynne}, {Amblard}, {Auld}, {Baes}, {Clements}, {Dariush}, {De
  Zotti}, {Dunne}, {Eales}, {Hopwood}, {Hoyos}, {Ibar}, {Jarvis}, {Maddox},
  {Micha{\l}owski}, {Pascale}, {Scott}, {Serjeant}, {Smith}, {Valiante}, \&
  {van der Werf}}]{Thacker2013}
{Thacker}, C., {Cooray}, {Smidt}, {et~al.} 2013, \apj, 768, 58


\bibitem[{{Viero} {et~al.}(2013){Viero}, {Wang}, {Zemcov}, {Addison},
  {Amblard}, {Arumugam}, {Aussel}, {B{\'e}thermin}, {Bock}, {Boselli}, {Buat},
  {Burgarella}, {Casey}, {Clements}, {Conley}, {Conversi}, {Cooray}, {De
  Zotti}, {Dowell}, {Farrah}, {Franceschini}, {Glenn}, {Griffin},
  {Hatziminaoglou}, {Heinis}, {Ibar}, {Ivison}, {Lagache}, {Levenson},
  {Marchetti}, {Marsden}, {Nguyen}, {O'Halloran}, {Oliver}, {Omont}, {Page},
  {Papageorgiou}, {Pearson}, {P{\'e}rez-Fournon}, {Pohlen}, {Rigopoulou},
  {Roseboom}, {Rowan-Robinson}, {Schulz}, {Scott}, {Seymour}, {Shupe}, {Smith},
  {Symeonidis}, {Vaccari}, {Valtchanov}, {Vieira}, {Wardlow}, \&
  {Xu}}]{Viero2013}
{Viero}, M.~P., {Wang}, L., {Zemcov}, M., {et~al.} 2013, \apj, 772, 77

\bibitem[Yue et~al. (2013)]{Yue2013}
Yue, B., Ferrara, A., Salvaterra, R., Xu, Y., \& Chen, X. 2013, MNRAS, 433, 1556–1566

\bibitem[Yue et~al. (2014)]{Yue2014}
Yue, B., Ferrara, A., Salvaterra, R., Xu, Y., \& Chen, X. 2014, MNRAS, 440, 1263-1273


\bibitem[{Zemcov} {et~al.}(2014)]{Zemcov2014}
{Zemcov}, M. and {Smidt}, J. and {Arai}, T. and {Bock}, J. and 
	{Cooray}, A., et~al. 2014, Science, 346, 732-735

\end{thebibliography}
\end{document}